\documentclass[reprint,superscriptaddress,prb]{revtex4}

\usepackage{graphicx}
\usepackage{epstopdf}
\usepackage{bbm}
\usepackage{amsmath}
\usepackage{eqnarray}
\usepackage{siunitx}
\usepackage{soul}
\usepackage{xcolor}

\begin{document}

\title{Electroluminescence of the  graphene 2D semi-metal}

\author{A. Schmitt}
\thanks{These two authors contributed equally to this work.}
\affiliation{Laboratoire de Physique de l'Ecole normale sup\'erieure, ENS, Universit\'e
PSL, CNRS, Sorbonne Universit\'e, Universit\'e Paris-Cit\'e, 24 rue Lhomond, 75005 Paris, France}
\author{L. Abou-Hamdan}
\thanks{These two authors contributed equally to this work.}
\affiliation{Institut Langevin, ESPCI Paris, PSL University, CNRS, 1 rue Jussieu, F-75005 Paris, France }
\affiliation{DOTA, ONERA, Université Paris-Saclay, F-91123 Palaiseau, France }
\author{M. Tharrault}
\affiliation{Laboratoire de Physique de l'Ecole normale sup\'erieure, ENS, Universit\'e
PSL, CNRS, Sorbonne Universit\'e, Universit\'e Paris-Cit\'e, 24 rue Lhomond, 75005 Paris, France}
\author{S. Rossetti}
\affiliation{Institut Langevin, ESPCI Paris, PSL University, CNRS, 1 rue Jussieu, F-75005 Paris, France }
\affiliation{DOTA, ONERA, Université Paris-Saclay, F-91123 Palaiseau, France }
\author{D. Mele}
\affiliation{Laboratoire de Physique de l'Ecole normale sup\'erieure, ENS, Universit\'e
PSL, CNRS, Sorbonne Universit\'e, Universit\'e Paris-Cit\'e, 24 rue Lhomond, 75005 Paris, France}
\affiliation{Univ. Lille, CNRS, Centrale Lille, Univ. Polytechnique Hauts-de-France, Junia-ISEN, UMR 8520-IEMN, F-59000 Lille, France.}
\author{R. Bretel}
\affiliation{Laboratoire de Physique de l'Ecole normale sup\'erieure, ENS, Universit\'e
PSL, CNRS, Sorbonne Universit\'e, Universit\'e Paris-Cit\'e, 24 rue Lhomond, 75005 Paris, France}
\author{A. Pierret}
\affiliation{Laboratoire de Physique de l'Ecole normale sup\'erieure, ENS, Universit\'e
PSL, CNRS, Sorbonne Universit\'e, Universit\'e Paris-Cit\'e, 24 rue Lhomond, 75005 Paris, France}
\author{M. Rosticher}
\affiliation{Laboratoire de Physique de l'Ecole normale sup\'erieure, ENS, Universit\'e
PSL, CNRS, Sorbonne Universit\'e, Universit\'e Paris-Cit\'e, 24 rue Lhomond, 75005 Paris, France}
\author{P. Morfin}
\affiliation{Laboratoire de Physique de l'Ecole normale sup\'erieure, ENS, Universit\'e
PSL, CNRS, Sorbonne Universit\'e, Universit\'e Paris-Cit\'e, 24 rue Lhomond, 75005 Paris, France}
\author{T. Taniguchi}
\affiliation{Advanced Materials Laboratory, National Institute for Materials Science, Tsukuba,
Ibaraki 305-0047,  Japan}
\author{K. Watanabe}
\affiliation{Advanced Materials Laboratory, National Institute for Materials Science, Tsukuba,
Ibaraki 305-0047,  Japan}
\author{J-M. Berroir}
\affiliation{Laboratoire de Physique de l'Ecole normale sup\'erieure, ENS, Universit\'e
PSL, CNRS, Sorbonne Universit\'e, Universit\'e Paris-Cit\'e, 24 rue Lhomond, 75005 Paris, France}
\author{G. Fève}
\affiliation{Laboratoire de Physique de l'Ecole normale sup\'erieure, ENS, Universit\'e
PSL, CNRS, Sorbonne Universit\'e, Universit\'e Paris-Cit\'e, 24 rue Lhomond, 75005 Paris, France}
\author{G. Ménard}
\affiliation{Laboratoire de Physique de l'Ecole normale sup\'erieure, ENS, Universit\'e
PSL, CNRS, Sorbonne Universit\'e, Universit\'e Paris-Cit\'e, 24 rue Lhomond, 75005 Paris, France}
\author{B. Plaçais}
\affiliation{Laboratoire de Physique de l'Ecole normale sup\'erieure, ENS, Universit\'e
PSL, CNRS, Sorbonne Universit\'e, Universit\'e Paris-Cit\'e, 24 rue Lhomond, 75005 Paris, France}
\author{C. Voisin}
\affiliation{Laboratoire de Physique de l'Ecole normale sup\'erieure, ENS, Universit\'e
PSL, CNRS, Sorbonne Universit\'e, Universit\'e Paris-Cit\'e, 24 rue Lhomond, 75005 Paris, France}
\author{J-P. Hugonin}
\affiliation{Laboratoire Charles Fabry, IOGS, Université Paris-Saclay, F-91123 Palaiseau, France}
\author{J-J. Greffet}
\affiliation{Laboratoire Charles Fabry, IOGS, Université Paris-Saclay, F-91123 Palaiseau, France}
\author{P. Bouchon}
\affiliation{DOTA, ONERA, Université Paris-Saclay, F-91123 Palaiseau, France }
\author{Y. De Wilde}
\affiliation{Institut Langevin, ESPCI Paris, PSL University, CNRS, 1 rue Jussieu, F-75005 Paris, France }
\author{E. Baudin}
\affiliation{Laboratoire de Physique de l'Ecole normale sup\'erieure, ENS, Universit\'e
PSL, CNRS, Sorbonne Universit\'e, Universit\'e Paris-Cit\'e, 24 rue Lhomond, 75005 Paris, France}

\begin{abstract}

Electroluminescence, a non-thermal radiative process, is ubiquitous in semi-conductors and insulators but fundamentally precluded in metals. We show here that this restriction can be circumvented in high-quality graphene.
By investigating the radiative emission of semi-metallic graphene field-effect transistors over a broad spectral range, spanning the near- and mid-infrared, we demonstrate direct far-field electroluminescence from hBN-encapsulated graphene in the mid-infrared under large bias in ambient conditions. 
Through a series of test experiments ruling out its incandescence origin, we determine that the electroluminescent signal results from the electrical pumping produced by interband tunneling. We show that the mid-infrared electroluminescence is spectrally shaped by a natural quarter-wave resonance of the heterostructure. This work invites a reassessment of the use of metals and semi-metals as non-equilibrium light emitters, and the exploration of their intriguing specificities in terms of carrier injection and relaxation, as well as emission tunability and switching speed.

\end{abstract}

\maketitle

\section{Introduction}\label{s:intro}

Today, electroluminescent devices represent the major part of lighting and display devices and are building blocks of the global information and telecommunication network. \cite{akasaki2015nobel} Electroluminescence is a non-thermal phenomenon by which a material emits light in response to an electrical current. \cite{rosencher2002optoelectronics}
In solids, it is prevalent in insulators and semi-conductors \cite{piper1955theory} and more recently in organic materials and results from the radiative recombination of electrons and holes across the material's bandgap. 
Electroluminescence is precluded in metals, as their electron gas relaxes to a thermal state on a very short timescale due to their gapless electronic band structure. Concordantly, they can only exhibit incandescent emission (a thermal radiative process) from their hot Fermi sea. 
The question is still open for semi-metals, as they share similarities with both metals (absence of bandgap) and semi-conductors (presence of a valence and a conduction band), and it is addressed in the present paper.

One may naturally wonder whether an exotic regime could arise beyond thermal equilibrium, in which a metal would exhibit luminescence. 
A simple way to reach an out-of-equilibrium regime is by pumping the material using an intense and very short light pulse. The large energy brought by individual photons allows charge carriers to reach high energy levels and, under strong pumping, a non-equilibrium state may survive long enough before thermalization for a transient photoluminescence signal to be observed. This scheme has been achieved in various metals,  more specifically via the selective excitation of confined plasmons.\cite{WOS:000234337900090} Continuous-wave photoluminescence of metals is also possible, but with an extremely low efficiency due to the competing relaxation processes\cite{mooradian1969photoluminescence}.
Electroluminescence is even more elusive as electrical excitation lacks energy selectivity\footnote{When metals and semi-metals are used as an injector in a tunneling junction, they can lead to electroluminescence via inelastic tunneling; such a process, however, is extrinsic to the electronic state of the material.}.

The question of electroluminescence in semi-metals can now be revisited thanks to the emergence of high-mobility graphene transistors behaving as model semi-metallic devices. In particular, recent advancements in the fabrication of intrinsic hexagonal Boron Nitride (hBN)-encapsulated graphene field-effect transistors have enabled reaching intrinsic material limits\cite{Betz2012nphys,Graham2012nphys}. This opens new exotic transport regimes \cite{yang2018graphene,Kumar2017nphys,Bandurin2016science,schmitt2023mesoscopic,Berdyugin2022science}, in which an almost perfect decoupling of the electron gas from its surroundings can be achieved at room temperature and ambient pressure. Furthermore, the ballistic transport of electrons over micrometric scales triggers new interband conduction regimes (such as Zener-Klein tunneling), thereby driving the Fermi sea far out of equilibrium. \cite{yang2018graphene,baudin2020hyperbolic,schmitt2023mesoscopic}

Graphene has been extensively studied as a novel broadband incandescent emitter due to its atomic thickness and ultra-small heat capacity.\cite{kim2018ultrafast,luo2019graphene,berciaud2010prl}
As a metal, graphene also exhibits transient hot photoluminescence\cite{lui2010ultrafast,kim2021mid}, transient athermal electronic distribution\cite{gierz2013snapshots, johannsen2013direct}, and inelastic tunneling electroluminescence \cite{beams2014electroluminescence, kuzmina2021resonant}. 
However, to date, the prerequisite for electroluminescence has yet to be reported, namely the direct observation of an out-of-equilibrium electron-hole plasma obtained under electrical pumping. 
Here, we report on the direct observation of the continuous  far-field electroluminescence from graphene transistors in the mid-infrared (MIR) originating from the electron-hole pairs injected by the interband Zener-Klein tunneling mechanism under high bias.

\section{\label{sec:levelhist-1}the electroluminescence mechanism in graphene }

\begin{figure}
\begin{centering}
\includegraphics[width=14cm]{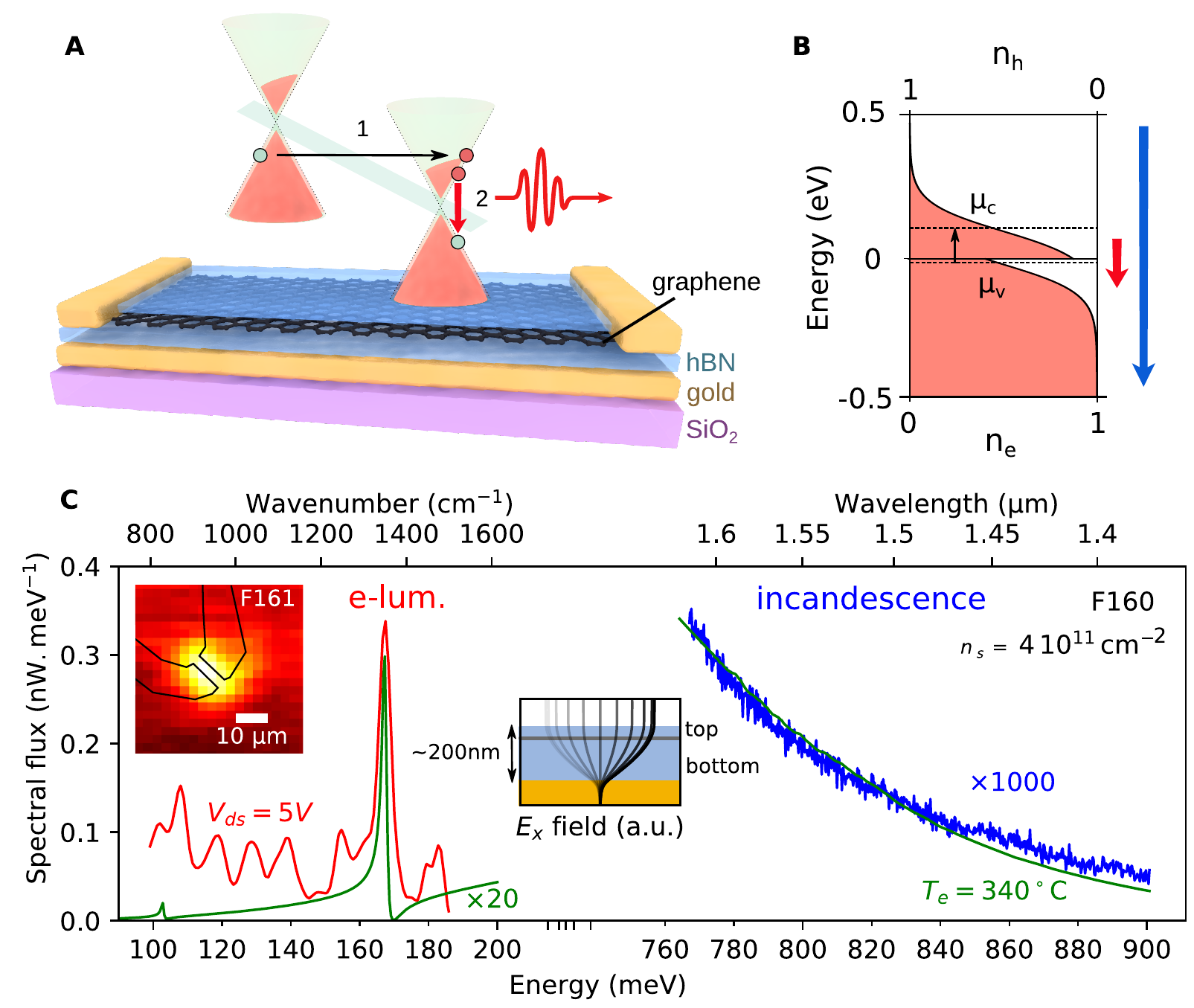}
\par\end{centering}
\caption{\label{fig:Overview} \textbf{Incandescence and electroluminescence
in a high-mobility graphene transistor.} (A) Step 1: At large bias,
Zener-Klein tunneling of electrons occurs between the valence and
the conduction band. Step 2: Electron and hole pockets develop in the conduction and valence bands. Direct interband electron-hole recombination at low
energy (red arrow) leads to MIR electroluminescence, whereas high-energy recombination
(blue arrow in panel B) of thermal carriers leads to NIR incandescence. 
(B) Corresponding electron and hole distributions under electrical pumping.
(C) F160 NIR to MIR spectral flux (blue and red curves, respectively) under a $5\, \mathrm{V}$ drain-source bias. The green curves are the graphene computed spectral flux at $340{\rm ^{\circ}C}$ using the computed spectral emissivity of graphene alone (see supplementary information, section~VI.B.2). (Left inset) Spectrally-integrated MIR emission map of the electrically biased F161 transistor (see supplementary information, section~II.B). (Middle inset) Transverse electric field distribution at resonance.} 
\end{figure}

Electroluminescence requires an ambipolar electrical injection
mechanism to drive the electronic distribution out of equilibrium.
\cite{rosencher2002optoelectronics} In semi-conductor devices, such as light-emitting diodes (LEDs), electrical injection is achieved by the ambipolar current passing through a PN junction. We use high-mobility graphene field-effect transistors (HGFETs), which allow for versatile control of channel doping and electrical pumping (see below). In contrast to LEDs, HGFETs are unipolar, but, under
large electrical bias, they sustain a large interband current due to the bulk Zener-Klein tunneling of electrons from the valence band to the conduction band. \cite{yang2018graphene} As depicted schematically in Fig.~\ref{fig:Overview}A, this charge carrier injection process builds an out-of-equilibrium electronic distribution, which instigates electroluminescence. This stationary out-of-equilibrium electronic distribution, 
which is illustrated in Fig.\ref{fig:Overview}B, is made possible in graphene for two reasons. (i) The presence of a relaxation bottleneck at the Dirac point separating the valence and conduction bands. (ii) The large hBN and graphene optical phonon energies ($\sim 170\,\rm{meV}$), which ultimately bound the out-of-equilibrium electronic distribution. Therefore, we can expect electroluminescence signals from HGFETs in the $50-200\,\rm{meV}$ MIR spectral range, while incandescence is expected in the near-infrared (NIR) to the visible spectral range.

The MIR emission map of a high-mobility graphene transistor under electrical bias is shown in Fig.~\ref{fig:Overview}C (inset). As expected, the measured MIR emission originates from the transistor's graphene channel. To investigate the nature of this emission we measured the MIR and NIR emission spectra (Fig.~\ref{fig:Overview}C) of a high-mobility graphene FET equipped with a gold backgate (\textit{F160}, $\mu=13\,\rm{m^2.V^{-1}.s^{-1}}$) under electrical bias at ambient conditions (see SI sections II and III for experimental details). The two spectra are drastically different; namely, the NIR spectrum ($0.76-0.9$~eV) follows the behavior of a typical Planckian gray body at $340{\rm ^{\circ}C}$ (green line), whereas the MIR spectrum ($0.1-0.2$~eV) exhibits a solitary resonance peak at $1348$~cm$^{-1}$($\equiv 167$~meV), corresponding to a quarter-wave resonance of the hBN/graphene/hBN/gold heterostructure (see below). Importantly, the MIR peak emission exceeds that of the thermal incandescent emission (green curve) by a factor of 20, the latter being calculated from the incandescence temperature determined in the NIR spectrum.

We detail now the mechanism responsible for the electroluminescence peak. The spectral shaping of the emission is dominated by the light-matter coupling
strength modulation imposed by the electromagnetic environment. \cite{wojszvzyk2021incandescent,Ghirardini20212Dmat} 
For the transistors considered here, their highly reflecting gold backgate ($160\,\rm{nm}$ below the graphene channel) decouples graphene from the far-field over the entire MIR spectral range under study, except
near the resonances of the refractive index $n$ of its direct environment, namely, the hBN substrate. Maximum coupling of the graphene layer's emission to the far-field occurs near the quarter-wave resonance of the heterostructure, \emph{i.e.}, at a wavelength $\lambda = 4n(\lambda)\, d$, where $d$ is the thickness of the hBN layer separating the graphene layer from the gold backgate. 
As an example, for the F160 transistor (Fig.~\ref{fig:Overview}C), in which the total hBN thickness is 0.205~$\mu$m, this condition is satisfied at $\lambda = 7.41$~$\mu$m ($\equiv 1349$~cm$^{-1}$), at which $n = 9.1$. Taking into account oblique incidence and the top air/hBN interface only results in a slight redshift, down to a resonance wavelength of $7.43$~$\mu$m ($\equiv 1348$~cm$^{-1}$). For F160, this single emission peak is observed both experimentally (Fig.~\ref{fig:Overview}C, red curve) and from incandescence numerical calculations (Fig.~\ref{fig:Overview}C, green curve). This quarter-wave mediated electroluminescence is robustly observed in all the investigated samples (see Supplementary Figure SI-13). 

We note that the use of a gold backgate is key to unambiguously observing electroluminescence in graphene field-effect transistors. Firstly, it allows direct coupling between graphene's electron gas and the far-field. Secondly, it minimizes the incandescent thermal background of the heated substrate. Finally, it relegates the emission peak to frequencies below hBN's \emph{Reststrahlen} band ($1360-1620 \, \rm{cm^{-1}}$) thereby precluding the possibility that emission could occur due to scattering of the near-field modes confined in hBN. 
In principle, graphene's plasmonic branch could also contribute to the signal via resonant Rayleigh scattering by defects into the quarter-wave mode. However, the  extremely short plasmon wavelength\cite{Brar2014nanol,principi2017super} $\lambda \sim 25\, \rm{nm}$ at $\omega=1348\,\rm{cm^{-1}}$ makes this deep plasmon out-of-reach, even in fully optimized dedicated scattering-scanning near-field optical microscopy measurements ($\lambda_{\rm{min}} \sim  100\, \rm{nm}$).\cite{dai2014tunable, dai2015graphene, woessner2015highly}

\section{Ruling out incandescence}

\begin{figure}
\begin{centering}
\includegraphics[width=8cm]{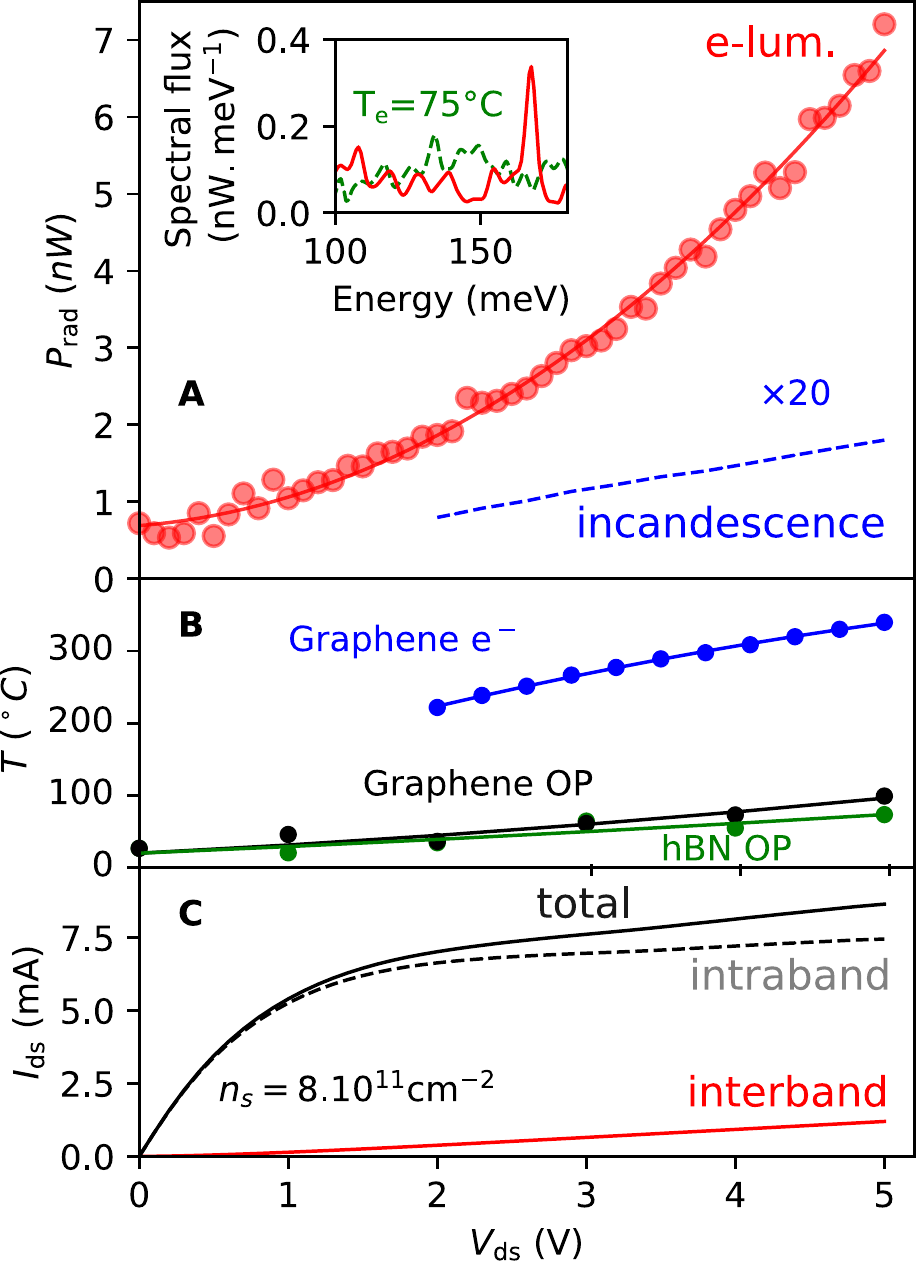}
\par\end{centering}
\caption{\label{fig:Nature-of-EL}\textbf{Nature of the light emission from graphene in the high-mobility transistor F160 as a function of bias.}
 (A) Spectrally-integrated MIR radiated power in the $[800, 1500]\, \rm{cm^{-1}}$ band (red circles). The dashed blue curve represents the radiated power expected from graphene's incandescence when heated at the measured electron gas temperature shown in panel B.  (Inset) Comparison of the MIR spectra of F160 under bias (red curve) and when heated (green curve) to the temperature of hBN's optical phonons (panel B). (B) Temperature of the optically-active quasi-particles of the heterostructure: the optical phonons (OP) of graphene and hBN, and graphene's electrons. (C) F160's current-to-voltage response. Interband current (red line) is deduced from the formula given in the Supplementary information. 
}
\end{figure}

Having thoroughly illustrated the near- to mid-infrared spectral features of the transistor's emission, we turn now to the behavior of its $[800-1500]\, \rm{cm^{-1}}$ spectrally-integrated MIR emission as a function of increasing bias (Fig.~\ref{fig:Nature-of-EL}A). It is instructive to compare the actual emitted power with that which is anticipated from incandescence, that is, $P_{\rm{rad,th}}(V_{ds}) \simeq (2.11\, \rm{nW}) \times \left( \frac{1}{\exp (\hbar \omega / k_B T_e (V_{ds})) -1}-\frac{1}{\exp (\hbar \omega / k_B T_0 ) -1}\right)$ (Fig.~\ref{fig:Nature-of-EL}A, blue curve). This radiative power is calculated by spectrally integrating the product of the device's emissivity and the blackbody spectral radiance at the measured electronic temperature (determined from the graphene layer's NIR incandescence, Fig.~\ref{fig:Nature-of-EL}B, blue curve, see methods in SI section III). In the above expression, $\hbar \omega$ is the quarter-wave resonance energy, $T_e$ is the measured electronic temperature, $T_0=293\,\rm{K}$ is the environment temperature, and $k_B$ is the Boltzmann constant. The actual emission follows a super-linear trend, while the anticipated incandescent emission follows a quasi-linear trend with bias.

Spectrally-shaped incandescence can easily be mistaken for electroluminescence\cite{essig2010phonon} so that the comparison of emission powers is crucially important to exclude incandescence. 
In our case, the power of the measured spectrally integrated MIR emission is found to be at least 70 times larger than that of the computed incandescence of graphene's hot electrons.

We also rule out incandescence from all other optically active quasi-particles within the heterostructure, namely, the optical phonons of graphene and hBN. Their measured temperature ($T_{\textrm{OP,hBN}}\leq 75^\circ \rm{C}$, $T_{\textrm{OP,Gr}}\leq 95^\circ \rm{C}$, obtained via Raman Stokes-anti-Stokes thermometry, see SI section IV) lies well below the electronic temperature (see Fig.~\ref{fig:Nature-of-EL}B). Note that, since the emittance of hBN is 47 times larger than that of graphene at the quarter-wave resonance ($1348$~cm$^{-1}$), even a tiny temperature increase of hBN to 75$^{\circ}$C could potentially mimic the observed radiation spectrum from the transistor. Therefore, we compare the emission spectrum of the electrically-biased transistor to its emission when externally heated to the temperature of hBN's optical phonons (Fig.~\ref{fig:Nature-of-EL}A, inset). This comparison shows that thermal emission at the quarter-wave resonance is at most one-fourth of the radiated power under bias, thus excluding any possible thermal origin of the observed MIR signal. 

To quantitatively analyze the out-of-equilibrium electron gas emission of the transistor under bias, we utilize the Roosbroeck-Shockley theory of luminescence within the framework of the local Kirchhoff law. \cite{greffet2018light} Within this picture, the radiated power at energy $\hbar \omega$ scales as the product of electron and hole densities $n_e n_h$, which takes the form of a modified Bose-Einstein distribution given by $n_e n_h=\left( \exp\left(\frac{\hbar\omega - \Delta \mu}{k_{B}T}\right)-1\right)^{-1}$ at finite temperature $T$ and electron-hole chemical potential imbalance $\Delta \mu$ produced by electrical pumping. \cite{rosencher2002optoelectronics} 

Due to spectral shaping by the quarter-wave resonance, we apply this scaling law solely on the spectral flux obtained  at the $\omega = 1348\, \rm{cm^{-1}}$ resonance peak of Fig.~\ref{fig:Overview}C.
This worst-case-scenario calculation gives a 20-fold increase in electroluminescent radiated power with respect to incandescence. It also enables the direct quantification of the chemical potential imbalance (see SI section VI.B), yielding $\Delta \mu \geq 128 \, \rm{meV}$, \emph{i.e.}, exceeding the thermal energy  $k_B T\simeq 54 \, \rm{meV}$. The condition $\Delta \mu > k_B T$ is commonly accepted as the boundary for electroluminescent operation in LEDs\cite{greffet2018light}.  Regarded as incandescence, this would coincide with the emission power that would be reached for a graphene layer heated to $2600\,\mathrm{K}$, \emph{i.e.}, in excess of its sublimation temperature\cite{huang2009situ} ($\sim 2300\,\rm{K}$).  This simple consideration assumes that emittance is constant when the device is biased or heated, a property that we verify in the Supplementary Information Section VI.B.3. 

As a last step, we highlight the contrasting behaviors of the total current (Fig.~\ref{fig:Nature-of-EL}C, black curve) and the radiated power (Fig.~\ref{fig:Nature-of-EL}A)  with increasing bias. In particular, the radiated power evolves super-linearly, while the total current increases sub-linearly. This behavior deviates from that of conventional LEDs in which the radiated power is proportional to the current flowing through the device. Such striking behavior is thus singular to electroluminescence in semi-metals.

\begin{figure}
\begin{centering}
\includegraphics[width=16cm]{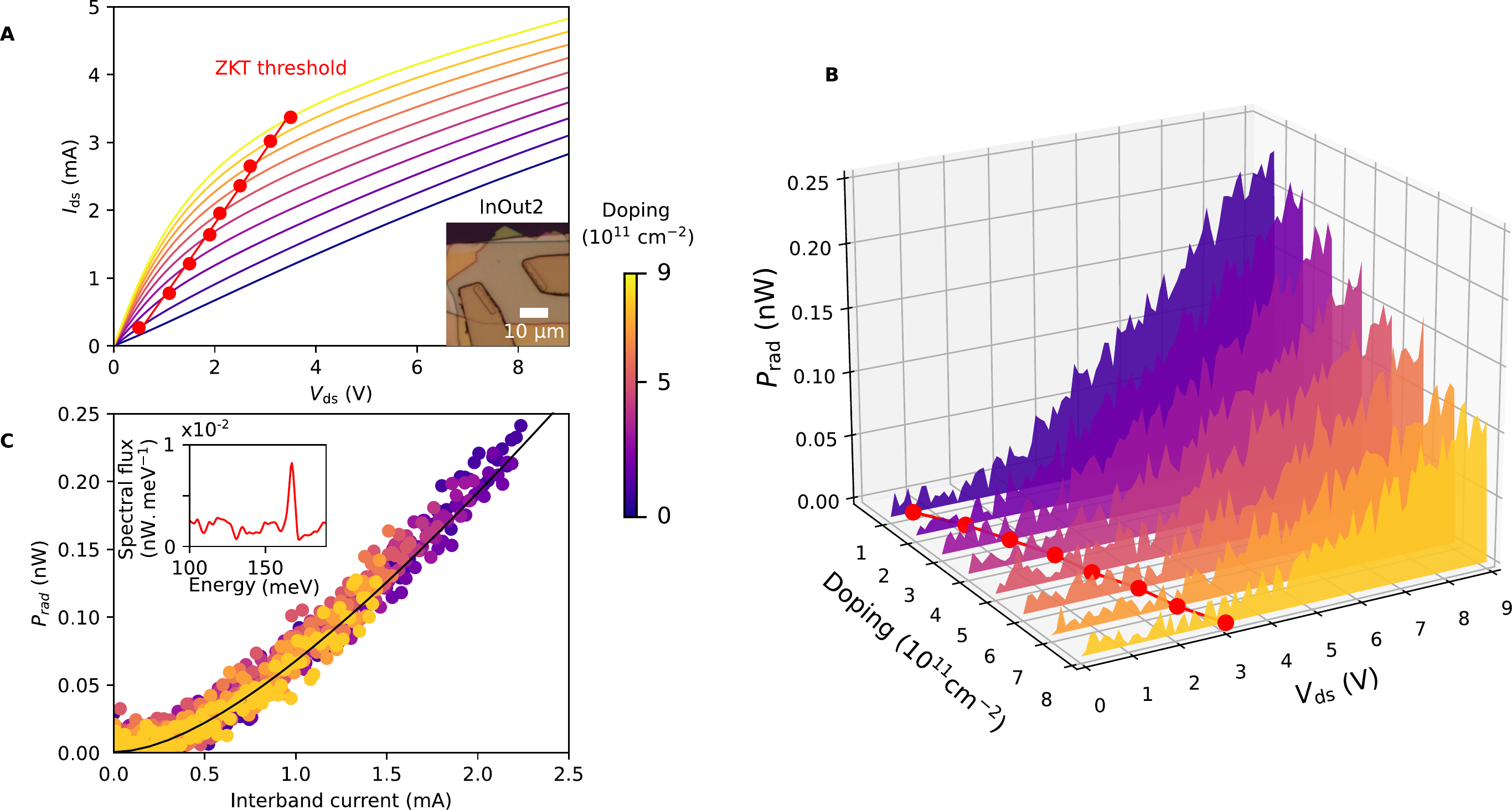}
\par\end{centering}
\caption{\label{fig:electrical-injection}\textbf{Electrical injection in
a long high-mobility graphene transistor.} (A) Characteristic current-to-voltage
response of the InOut2 transistor (shown in the inset). The color scale indicates the doping level. (B) Corresponding $[800, 1500]\, \rm{cm^{-1}}$ spectrally-integrated MIR radiative power as a function of applied
bias. The Zener-Klein tunneling (ZKT) threshold voltages are superimposed as red dots. (C)
Radiated power as a function of interband current. The carrier balance model is represented as a black line.  Inset: MIR spectrum under bias. 
}
\end{figure}

\section{Electrical injection in graphene}

Since the interband current is the sole instigator of electroluminescence, it is necessary to isolate its contribution to the transistor's total current as in Fig.~\ref{fig:Nature-of-EL}C. The presence of an interband current is unambiguously signaled here by the fact that, at large bias, the total current increases independently of the number of charge carriers (see Fig.~\ref{fig:electrical-injection}A). This interband current arises due to the Zener-Klein tunneling at large bias. \cite{yang2018graphene} 
The onset of this Zener-Klein tunneling regime invariably occurs at a threshold voltage which is set by Pauli blocking on the elementary tunneling events between the valence and the conduction bands (see Fig. \ref{fig:Overview}A). As a consequence, tunneling begins at a finite threshold bias $V_{th}=2E_{F}/e\times L/l_{ZK},$ where
$E_{F}=\hbar v_{F}\sqrt{\pi n_{s}}$ is the Fermi energy (with Fermi velocity $v_{F}$ and doping concentration $n_{s}$), $L$ is
the device length, and $l_{ZK}$ is the characteristic length
of Zener-Klein tunneling\cite{yang2018graphene}. In what follows, we consider a transistor, hereafter referred to as InOut2, whose long graphene channel (see Fig.~\ref{fig:electrical-injection}A, inset) enables large threshold voltages. In practice, the Zener-Klein tunneling threshold voltage is determined by
fitting the transistor's IV response curve with a phenomenological formula (see SI section V). The resulting thresholds for InOut2 are indicated by red circles in Fig.~\ref{fig:electrical-injection}A which clearly illustrate their doping
dependence.

The direct role of interband Zener-Klein tunneling as the initiator of electroluminescence is confirmed by the measured spectrally-integrated MIR radiated power of InOut2 (Fig.~\ref{fig:electrical-injection}B), which shows that the emission invariably starts at the tunneling threshold. To interpret the radiative emission in more detail, we further consider the spectrally-integrated radiative power as a function of interband current which presents a linear evolution above a characteristic threshold current $\sim 0.5\,\rm{mA}$  
(see Fig.~\ref{fig:electrical-injection}C). This behavior is reminiscent of non-radiative recombination pathways in common LEDs and can be explained, in the low doping limit, by the balance between the interband electron-hole pair creation rate ($I_{\rm{inter}}/2e$) and the electron-hole pair non-radiative ($A_{\rm{nr}} N$) and radiative ($B_{\rm{rad}} N^2$) rates, where $N$ is the excess carrier number\cite{rosencher2002optoelectronics}. As the observed radiative power $P_{\rm{rad}}$ is proportional to the radiative term, we deduce that $P_{\rm{rad}} \propto  \left( \sqrt{I_{\rm{inter}}+ I_0} - \sqrt{I_0} \right)^2,$ with $I_0 = e A^2_{\rm{nr}}/B_{\rm{rad}}.$ This model is shown in Fig.~\ref{fig:electrical-injection}C (black curve) and correctly captures the experimental behavior.

\section{Conclusion}

In closing, we have unequivocally showed that graphene, a 2D semi-metal, exhibits DC electroluminescence in the MIR. 
This is made possible at large electrical bias due to a Zener-Klein tunneling current sustaining an out-of-equilibrium electronic distribution accross the Dirac point. In addition to its significance from the fundamental point of view, the electroluminescence of graphene has interesting implications for the development of novel light sources. Indeed, the chemical potential imbalance reported here ($\Delta \mu \gtrsim 128\, \rm{meV}$) approaches the emission energy of $167\, \rm{meV}$, above which the optical amplification regime is reached. Interestingly, this extreme hot electron distribution, much more pronounced than the one associated to metals photoluminescence, is created by a mere electric field and could be highly relevant for hot electron 
photocatalysis \cite{Kim2017aom}. The current limiting factors to far-field electroluminescence are the competing non-radiative \cite{Winzer2012prb} and near-field radiative\cite{tielrooij2018out,yang2018graphene,baudin2020hyperbolic,Pogna2021acsnano} relaxation processes. 
Finally, we highlight the fact that
 the gapless band structure of 2D semi-metals enables electroluminescence at arbitrary photon energy. In principle, by engineering the out-of-plane coupling of the graphene layer with its dielectric surroundings, it should be possible then to evoke electroluminescence over a large span of wavelengths \cite{Jago2015prb}. 

\section{Authors contributions}

AS, LAH, MT, PM, BP, YDW and EB conceived the experiments. AS conducted device fabrication and electrical measurements under the guidance of DM, MR and AP in the early developments. TT and KW provided the hBN crystals. 
LAH and SR made the MIR emission measurements with the help of AS.
SR and PB made the MIR reflectance measurements with the help of AS and RB.
AS performed the Raman thermometry measurements. 
MT preformed the NIR measurements.
AS, LAH, JJG, YDW, CV, BP, JMB, GF, GM and EB analyzed the data and developed the theoretical interpretation. 
JPH wrote the numerical codes. 
AS, LAH, YDW, and EB wrote the manuscript with contributions from the coauthors.

\section{Acknowledgments}

\noindent We thank Claire Li for preliminary measurements.

\noindent The research leading to these results has received partial funding from the European Union Horizon 2020 research and innovation program under grant agreement No.881603 "Graphene Core 3", and from the French  ANR-21-CE24-0025-01 "ELuSeM". This work has received support under the program “Investissements d’Avenir” launched by the French Government.

\section{Additional information}

\noindent Competing financial interests: The authors declare no competing financial interests.

\section{Data availability}

Data will be publicly available on Zenodo.

\bibliographystyle{unsrt}
\bibliography{ref}

\end{document}


\title{Electroluminescence of the graphene 2D semi-metal\\ (Supplementary Information)}

\author{A. Schmitt}
\thanks{These two authors contributed equally to this work.}
\affiliation{Laboratoire de Physique de l'Ecole normale sup\'erieure, ENS, Universit\'e
PSL, CNRS, Sorbonne Universit\'e, Universit\'e Paris-Cit\'e, 24 rue Lhomond, 75005 Paris, France}
\author{L. Abou-Hamdan}
\thanks{These two authors contributed equally to this work.}
\affiliation{Institut Langevin, ESPCI Paris, PSL University, CNRS, 1 rue Jussieu, F-75005 Paris, France }
\affiliation{DOTA, ONERA, Université Paris-Saclay, F-91123 Palaiseau, France }
\author{M. Tharrault}
\affiliation{Laboratoire de Physique de l'Ecole normale sup\'erieure, ENS, Universit\'e
PSL, CNRS, Sorbonne Universit\'e, Universit\'e Paris-Cit\'e, 24 rue Lhomond, 75005 Paris, France}
\author{S. Rossetti}
\affiliation{Institut Langevin, ESPCI Paris, PSL University, CNRS, 1 rue Jussieu, F-75005 Paris, France }
\affiliation{DOTA, ONERA, Université Paris-Saclay, F-91123 Palaiseau, France }
\author{D. Mele}
\affiliation{Laboratoire de Physique de l'Ecole normale sup\'erieure, ENS, Universit\'e
PSL, CNRS, Sorbonne Universit\'e, Universit\'e Paris-Cit\'e, 24 rue Lhomond, 75005 Paris, France}
\affiliation{Univ. Lille, CNRS, Centrale Lille, Univ. Polytechnique Hauts-de-France, Junia-ISEN, UMR 8520-IEMN, F-59000 Lille, France.}
\author{R. Bretel}
\affiliation{Laboratoire de Physique de l'Ecole normale sup\'erieure, ENS, Universit\'e
PSL, CNRS, Sorbonne Universit\'e, Universit\'e Paris-Cit\'e, 24 rue Lhomond, 75005 Paris, France}
\author{A. Pierret}
\affiliation{Laboratoire de Physique de l'Ecole normale sup\'erieure, ENS, Universit\'e
PSL, CNRS, Sorbonne Universit\'e, Universit\'e Paris-Cit\'e, 24 rue Lhomond, 75005 Paris, France}
\author{M. Rosticher}
\affiliation{Laboratoire de Physique de l'Ecole normale sup\'erieure, ENS, Universit\'e
PSL, CNRS, Sorbonne Universit\'e, Universit\'e Paris-Cit\'e, 24 rue Lhomond, 75005 Paris, France}
\author{P. Morfin}
\affiliation{Laboratoire de Physique de l'Ecole normale sup\'erieure, ENS, Universit\'e
PSL, CNRS, Sorbonne Universit\'e, Universit\'e Paris-Cit\'e, 24 rue Lhomond, 75005 Paris, France}
\author{T. Taniguchi}
\affiliation{Advanced Materials Laboratory, National Institute for Materials Science, Tsukuba,
Ibaraki 305-0047,  Japan}
\author{K. Watanabe}
\affiliation{Advanced Materials Laboratory, National Institute for Materials Science, Tsukuba,
Ibaraki 305-0047,  Japan}
\author{J-M. Berroir}
\affiliation{Laboratoire de Physique de l'Ecole normale sup\'erieure, ENS, Universit\'e
PSL, CNRS, Sorbonne Universit\'e, Universit\'e Paris-Cit\'e, 24 rue Lhomond, 75005 Paris, France}
\author{G. Fève}
\affiliation{Laboratoire de Physique de l'Ecole normale sup\'erieure, ENS, Universit\'e
PSL, CNRS, Sorbonne Universit\'e, Universit\'e Paris-Cit\'e, 24 rue Lhomond, 75005 Paris, France}
\author{G. Ménard}
\affiliation{Laboratoire de Physique de l'Ecole normale sup\'erieure, ENS, Universit\'e
PSL, CNRS, Sorbonne Universit\'e, Universit\'e Paris-Cit\'e, 24 rue Lhomond, 75005 Paris, France}
\author{B. Plaçais}
\affiliation{Laboratoire de Physique de l'Ecole normale sup\'erieure, ENS, Universit\'e
PSL, CNRS, Sorbonne Universit\'e, Universit\'e Paris-Cit\'e, 24 rue Lhomond, 75005 Paris, France}
\author{C. Voisin}
\affiliation{Laboratoire de Physique de l'Ecole normale sup\'erieure, ENS, Universit\'e
PSL, CNRS, Sorbonne Universit\'e, Universit\'e Paris-Cit\'e, 24 rue Lhomond, 75005 Paris, France}
\author{J-P. Hugonin}
\affiliation{Laboratoire Charles Fabry, IOGS, Université Paris-Saclay, F-91123 Palaiseau, France}
\author{J-J. Greffet}
\affiliation{Laboratoire Charles Fabry, IOGS, Université Paris-Saclay, F-91123 Palaiseau, France}
\author{P. Bouchon}
\affiliation{DOTA, ONERA, Université Paris-Saclay, F-91123 Palaiseau, France }
\author{Y. De Wilde}
\affiliation{Institut Langevin, ESPCI Paris, PSL University, CNRS, 1 rue Jussieu, F-75005 Paris, France }
\author{E. Baudin}
\affiliation{Laboratoire de Physique de l'Ecole normale sup\'erieure, ENS, Universit\'e
PSL, CNRS, Sorbonne Universit\'e, Universit\'e Paris-Cit\'e, 24 rue Lhomond, 75005 Paris, France}



\date{\today}

\maketitle

\tableofcontents
\renewcommand{\thefigure}{SI-\arabic{figure}}

\section{Fabrication of high-mobility graphene transistors} 

The hexagonal-boron-nitride (hBN)-encapsulated graphene heterostructures are fabricated with the standard pick-up and stamping technique, using a polydimethylsiloxane(PDMS)/ polypropylene carbonate (PPC) stamp [\onlinecite{Yankowitz2019Nature}]. The gate electrode is first fabricated on a high-resistivity Si substrate covered by a 285~nm-thick $\text{SiO}_{\text{2}}$ layer. It consists of a pre-patterned gold pad (80~nm thick) designed by laser lithography and Cr/Au metalization. The hBN/Graphene/hBN heterostructure is then deposited on top of the backgate. This is followed by acetone cleaning of the stamp residues, Raman spatial mapping and AFM characterization of the stack. Graphene edge contacts are then created by means of laser lithography and reactive ion etching, securing low contact resistance $\lesssim 1\;\mathrm{k\Omega.\mu m}$. Finally, metallic contacts to the graphene channel are designed via a Cr/Au Joule evaporation.
The transistors' dimensions are maximized to get the highest optical signal, while their high mobility $\mu \gtrsim 10 \, \mathrm{m^{2}.V^{-1}.s^{-1}}$ at room temperature secures a moderate channel electric field $E= V/L \gtrsim 10^{5} \, \mathrm{V/m}$ for the threshold of mid-infrared electroluminescence.

\section{Mid-Infrared (MIR) Experimental Methods}

\subsection{Infrared spatial modulation spectroscopy (IR-SMS)}
\label{sec IRSMS}

\begin{figure}[t!]
\centering
\includegraphics[width=0.8\linewidth]{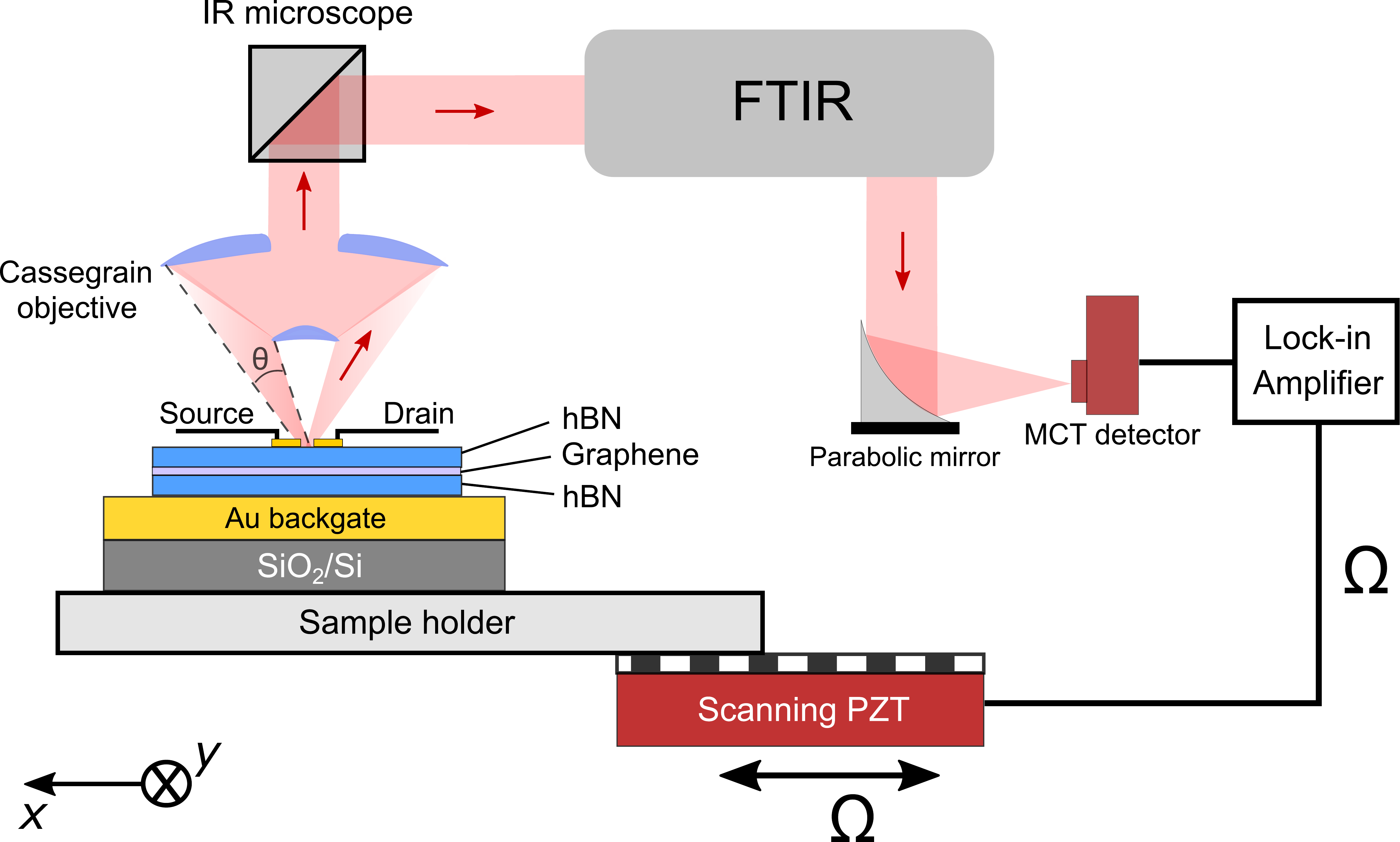}
\caption{Schematic illustration of the infrared spatial modulation spectroscopy (IR-SMS) setup [\onlinecite{li2018near, abou2021hybrid, abou2022transition}].}
\label{IRSMS}
\end{figure}

\begin{figure}[t!]
\centering
\includegraphics[scale=0.8]{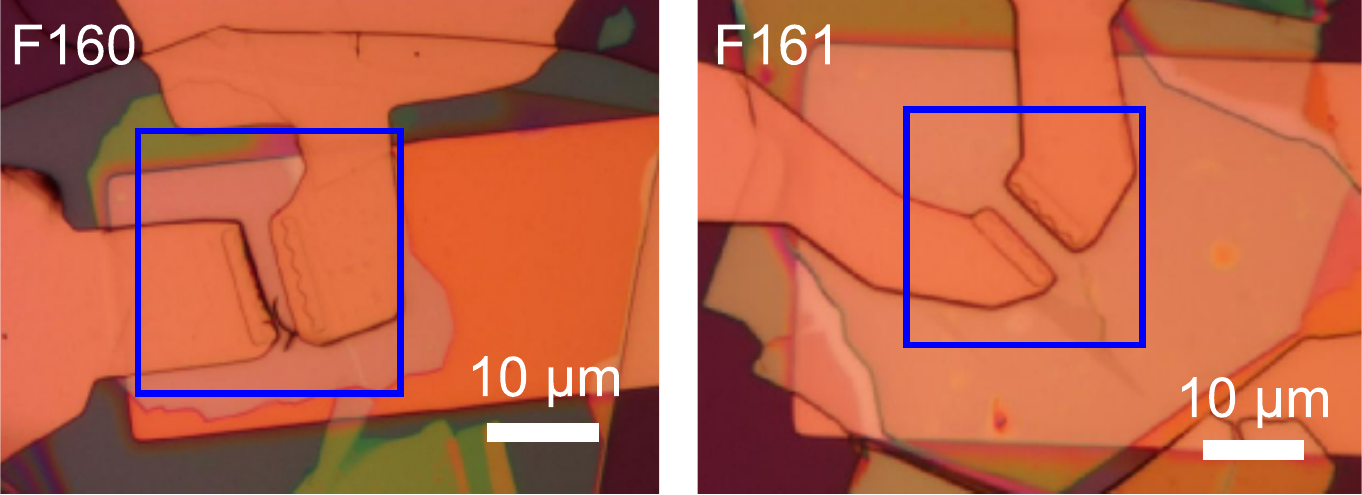}
\caption{ Microscope images of F160 and F161 with a blue square representing the estimated field of view of the MIR microscope.
}
\label{fig:F160F161_FOV}
\end{figure}

We measure the mid-infrared (MIR) electroluminescence spectra of the high-mobility graphene transistors via infrared spatial modulation spectroscopy (IR-SMS) [\onlinecite{li2018near, abou2021hybrid, abou2022transition}]. As sketched in Fig.~\ref{IRSMS}, the electroluminescent signal emitted by the transistor under electrical bias is collected using a Cassegrain objective with a large angular collection (numerical aperture, $\textrm{NA}=0.78$; angular collection interval, $\theta = 33-52^{\circ}$). The collected radiation is then guided through a step-scan operated Fourier transform infrared (FTIR) spectrometer before being refocused, via a parabolic mirror, onto a liquid-nitrogen-cooled mercury-cadmium-telluride (MCT) detector with a small active area $250 \times  250$~$\mu$m$^{2}$, which corresponds to a region of area $a =24.6\times 24.6$~$\mu$m$^{2}$ ($\times 10.16$ magnification) in the object plane (see Fig.~\ref{fig:F160F161_FOV}), produces a spatial selection.

The bare hBN/graphene/hBN channel between the gold electrodes of the transistor is \mt{optically conjugated with}{centered within} the active area of the MCT detector. The transistor is then modulated laterally using a piezoelectric translation stage at a frequency $\Omega$, \eb{by}{ over a distance} $\sim 25$~$\mu$m \eb{within}{ corresponding} to the microscope field of view. The optical signal reaching the active area of the detector, thus, varies between that of the graphene channel and \cv{}{that of }the Au electrode. The detected signal is demodulated at the frequency $\Omega$, allowing the measurement of a background-free electroluminescence spectrum. Alternatively, the FTIR can be bypassed so that the detected electroluminescent emission is integrated over the entire spectral bandwidth of the detector ($\lambda = 6 -14$~$\mu$m).

For the analysis of the thermal emission of our samples, electrical bias is suppressed and the sample holder is replaced by an electrically controlled hot-plate resistor, enabling uniform heating of the sample up to 573~K. The thermal emission measurements of the investigated samples are carried out at various temperatures as outlined above.\par
\begin{figure}[t!]
\centering
\includegraphics[width=\linewidth]{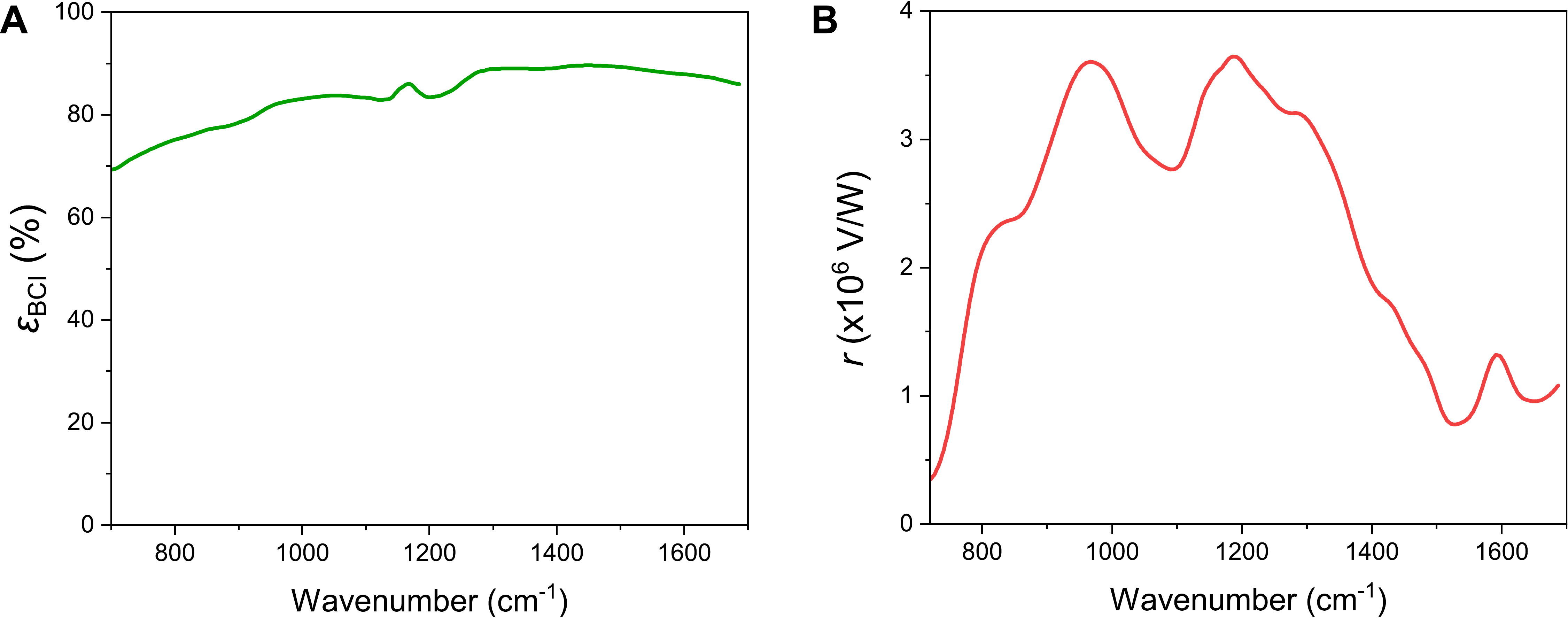}
\caption{(A) Measured mid-infrared emissivity of black carbon ink. (B) Spectral response function computed from Eq.~(\ref{eq:resp}).}
\label{response}
\end{figure}

The instrumental spectral response of the IR-SMS setup is determined by measuring the thermal emission of a reference sample at various temperatures (within the accessible range provided by the hot sample holder) and following a procedure similar to that outlined in Ref.~[\onlinecite{revercomb1988radiometric}]. The reference sample consists of a gold substrate that is partially covered with squares of black carbon ink (BCI). The squares are large enough to fully occupy the detector's field of view. Gold behaves as a mirror in the MIR range under study (reflectivity $\sim 99\, \%$), whereas BCI has a relatively high emissivity (see Fig.~\ref{response}A). The interface between a BCI square and the bare Au substrate is optically conjugated with the detector's field of view and a spectrum is acquired via IR-SMS. The detected signal at temperature $T$, $S(\nu, T)$ (where $\nu = 1/\lambda$ is the wavenumber in cm$^{-1}$), can then be written as follows.

\begin{equation}
\label{eq:sig}
S(\nu, T) = r(\nu) \big[\varepsilon_\text{BCI}M(\nu, T) + \rho_\text{BCI}M(\nu, T_{0})  + \rho_\text{Au}M(\nu, T_{0}) \big],
\end{equation}

\noindent
where $r(\nu)$ is the detector's spectral response function, $\varepsilon_\text{BCI}$ is the spectral emissivity of BCI, $\rho_\text{BCI}$ and $\rho_\text{Au}$ are the spectral reflectivities of BCI and gold, respectively, and $M = 4\pi h c^{2} \nu^{3}   \frac{1}{e^{hc  \nu / k_{B} T}   -1 }    \int_{0}^{a/2} \rm{d}x \rm{d}y   \int_{33^{\circ}}^{52^{\circ}} \sin{\theta} \cos{\theta}  \rm{d}\theta$ is the Planck blackbody spectral radiance in W/cm$^{-1}$ (integrated over the Cassegrain objective's angular collection interval and over half of the detector's field of view). The first term between brackets on the right-hand side of Eq.~(\ref{eq:sig}) corresponds to the spectral radiance of the \mt{}{heated} black carbon ink filling half of the detector's field of view, while the other two terms correspond to that of the room temperature ($T_{0}$) thermal radiation scattered off the sample. \par 
Using Eq.~(\ref{eq:sig}), the response function can be computed from the measured spectra of the reference sample at two different temperatures $T_{1}$ and $T_{2}$ ($T_{2}> T_{1}$) as follows

\begin{equation}
\label{eq:resp}
    r (\nu) = \frac{S (\nu, T_{2}) - S (\nu, T_{1})}{  \varepsilon_\text{BCI} [ M (\nu, T_{2}) - M (\nu, T_{1}) ]} .
\end{equation}

\begin{figure}[t!]
\centering
\includegraphics[width=0.8\linewidth]{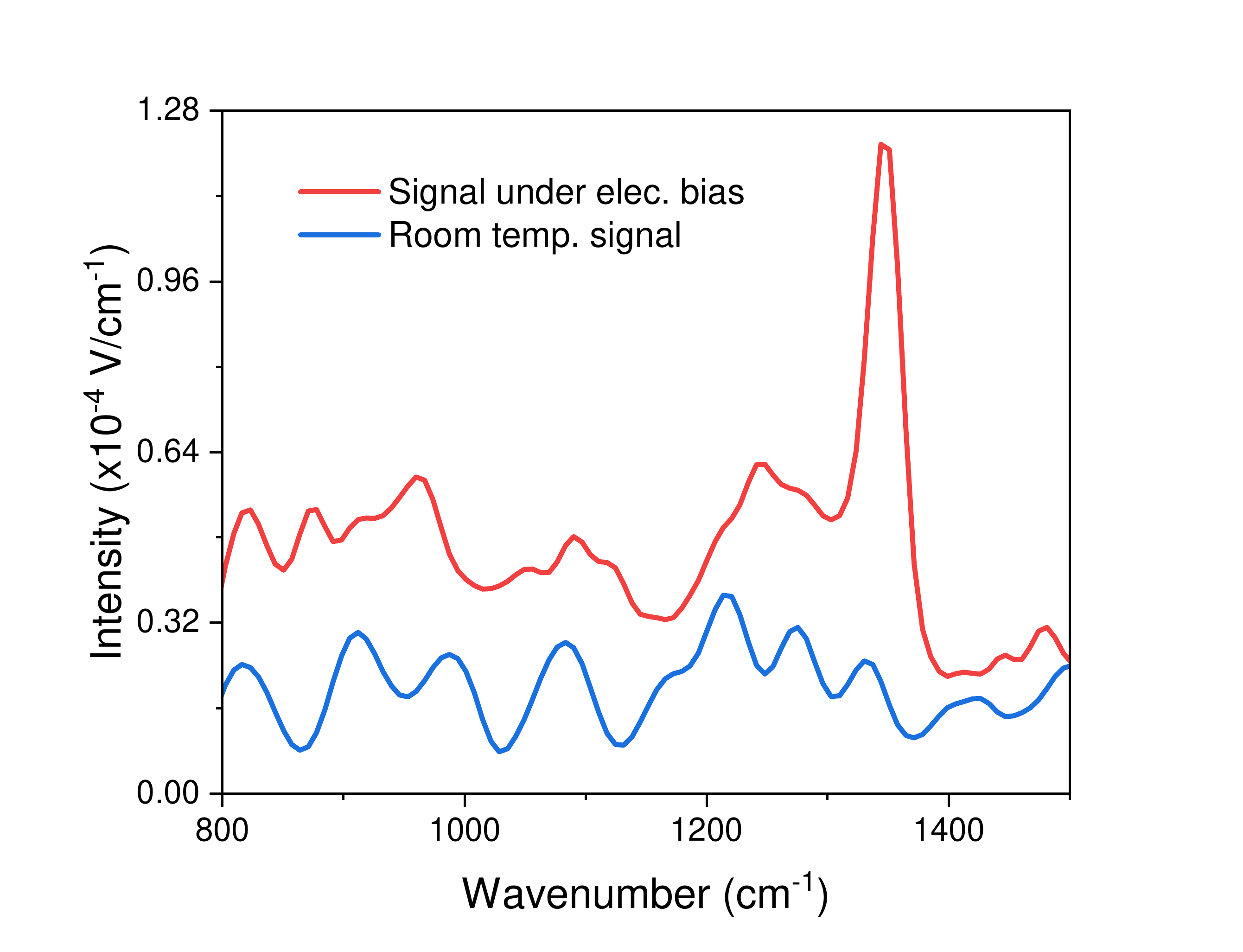}
\caption{Raw (unnormalized) IR-SMS spectra of F160 under electrical bias (red curve) and at room temperature (no electrical bias, blue curve).}
\label{elec bias vs room temp}
\end{figure}

\noindent
The resulting response function is plotted in Fig.~\ref{response}B. \par 
To eliminate the room temperature signal from the measured electroluminescence spectra, we measure the IR-SMS spectrum of the transistor under study at room temperature, which is then subtracted from the corresponding spectrum under electrical bias. The obtained spectrum is normalized by the response function (Eq.~(\ref{eq:resp})) to remove the instrumental dependence of the signal. The spectrum resulting from this normalization procedure corresponds to the spectral flux emitted by the transistor due to electrical pumping. The raw spectra of F160 under electrical bias and at room temperature with no electrical pumping are shown in Fig.~\ref{elec bias vs room temp}.

\subsection{Spectrally-integrated signal mapping}

\begin{figure}[ht!]
\centering
\includegraphics[width=0.8\linewidth]{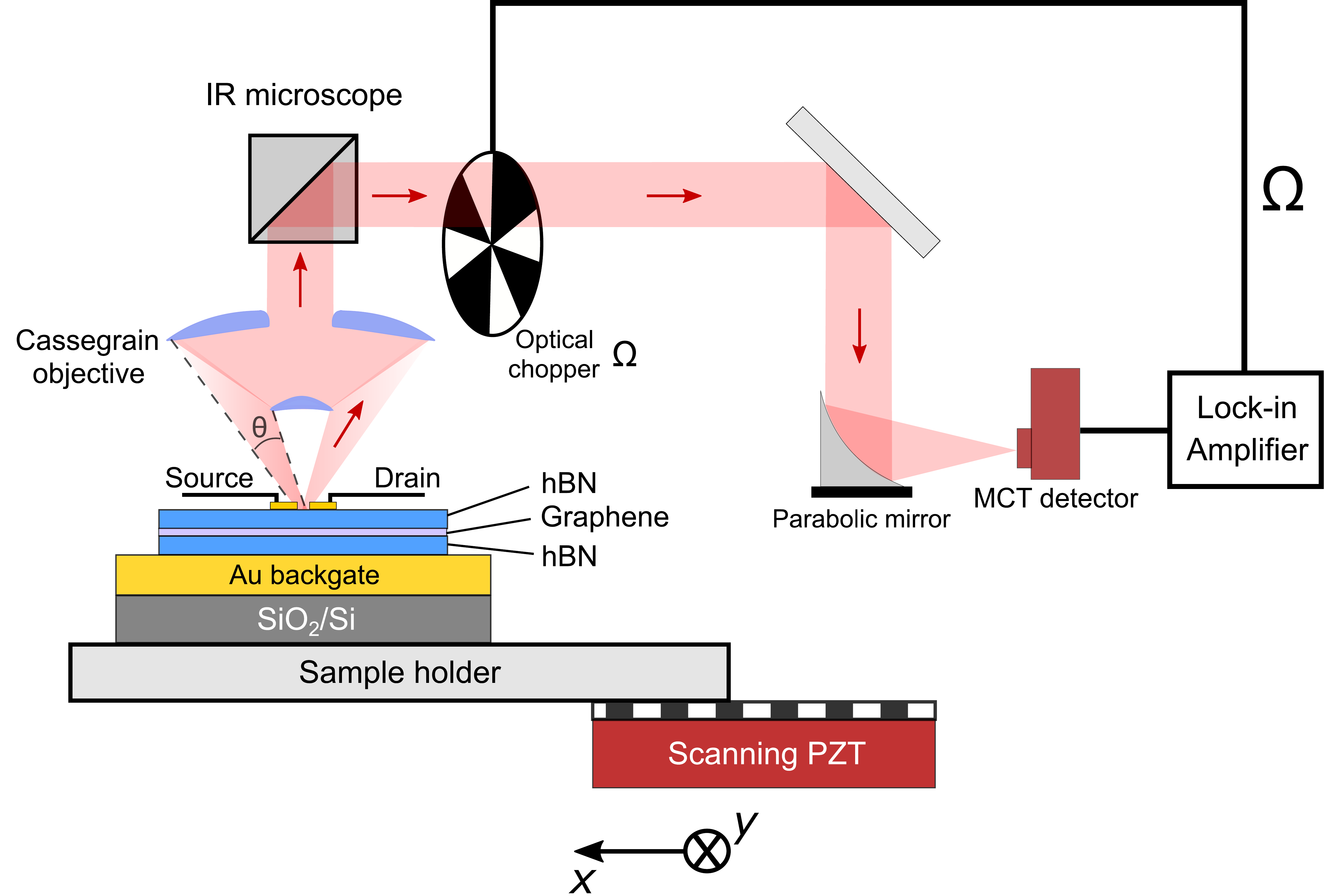}
\caption{Schematic illustration of the setup used to perform spatial scans of the spectrally-integrated infrared signal (integrated over the entire spectral range of the detector, $\lambda = 6 - 14$~$\mu$m).}
\label{chopper}
\end{figure}

To map the spatial distribution of the transistor's signal (as shown in the inset of Fig.~1C of the main text), we use the same setup described in subsection~\ref{sec IRSMS} except that the FTIR is bypassed and the lateral sample modulation is replaced by an optical chopper (see Fig.~\ref{chopper}). The transistor is raster scanned, using the piezoelectric translation stage, and the integrated signal (detected optical power within the full spectral range of the detector, $\lambda = 6 - 14$~$\mu$m) is recorded at each position. 

\begin{figure}[ht!]
\centering
\includegraphics[width=\linewidth]{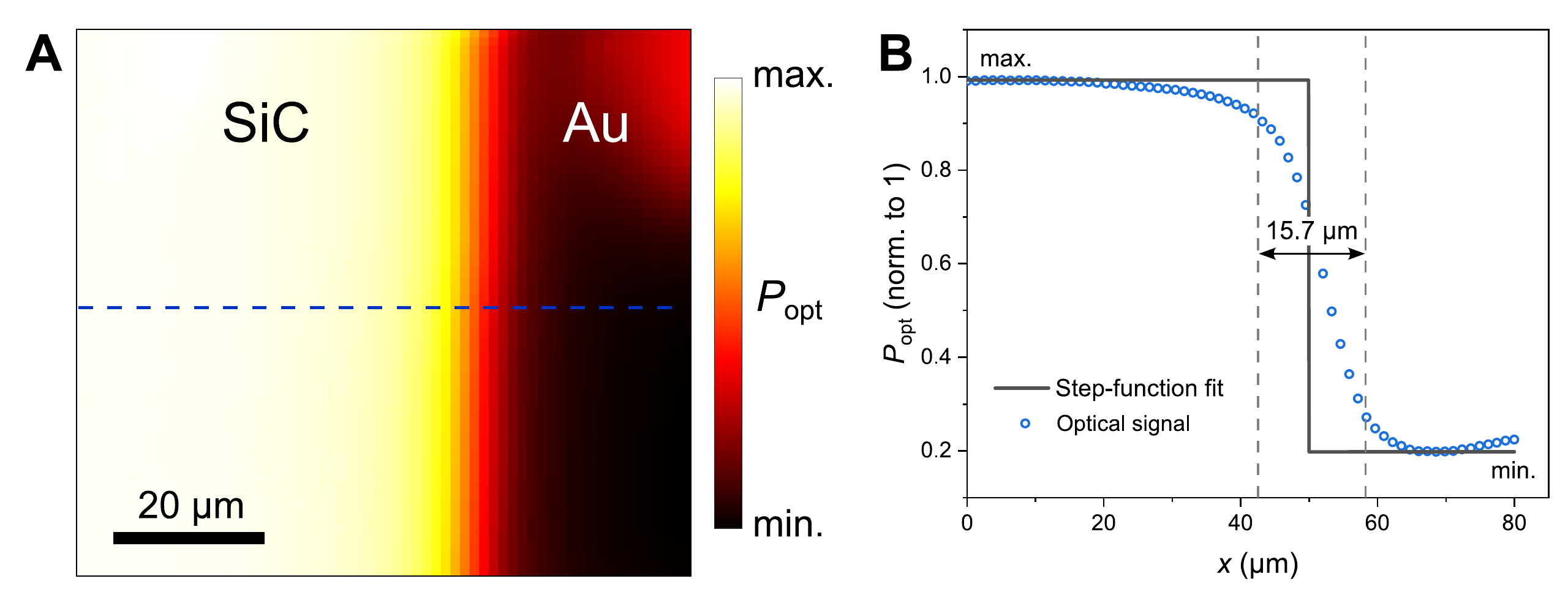}
\caption{(A) Image of the MIR optical signal ($P_{\textrm{opt}}$) measured at the interface between a SiC substrate and a large Au pattern. (B) Average horizontal line scan of panel (A). The two vertical lines correspond to the positions at which the signal reaches 10$\,\%$ and 90$\, \%$ of its maximal value. The imaging resolution is estimated to be 15.7~$\mu$m. }
\label{resolution}
\end{figure}

To estimate the imaging resolution, we imaged the interface between a SiC substrate and a large gold pattern (Fig.~\ref{resolution}A). The average line scan of the optical signal in the direction orthogonal to the interface (see Fig.~\ref{resolution}A, dashed blue line) follows a step-like behavior since SiC is highly emissive and Au is highly reflecting (see Fig.~\ref{resolution}B). We estimate from the line scan that the imaging resolution is around 15.7~$\mu$m. As such, the MIR emission spot observed in the inset of Fig.~1C of the main text appears larger than the graphene channel's physical dimensions.

\subsection{Absorptivity of a graphene transistor}
\label{MIR refl}

The MIR absorptivity of the graphene transistor F160 (Fig.~\ref{fig:hBNPeakMeas}A) is determined from its reflectivity, which is measured using a \emph{Bruker Hyperion} infrared microscope. A Cassegrain objective focuses a globar infrared source onto the sample and simultaneously collects the light scattered off the sample. FTIR spectroscopy is then performed on the light collected by the objective. A diaphragm is employed to limit the detector's field of view to only a $\sim$~($24\times 24$)-~$\mu$m$^{2}$ region of the sample (see Fig.~\ref{fig:F160F161_FOV}). A reference measurement is first performed on a large gold pattern on the sample, giving the intensity of incident light $I_{0}$. This is followed immediately by a measurement with the transistor's graphene channel in focus, giving an intensity $I$. The reflectivity is then obtained by taking the ratio of the two measured intensities, \emph{i.e.}, $R = I/I_{0}$. Since the transistor is on a gold backgate, the transistor's absorptivity is simply given by $A = 1-R$.

\section{Near-infrared blackbody spectroscopy}

\subsection{Experimental methods}
The temperature of graphene's electrons, under high drain-source bias, is obtained by probing its blackbody emission in the near-infrared (NIR) spectral range $1375 \, \mathrm{nm} - 1625 \, \mathrm{nm}$, using an InGaAs CCD camera.

\begin{figure}
\centering
\includegraphics[width=\linewidth]{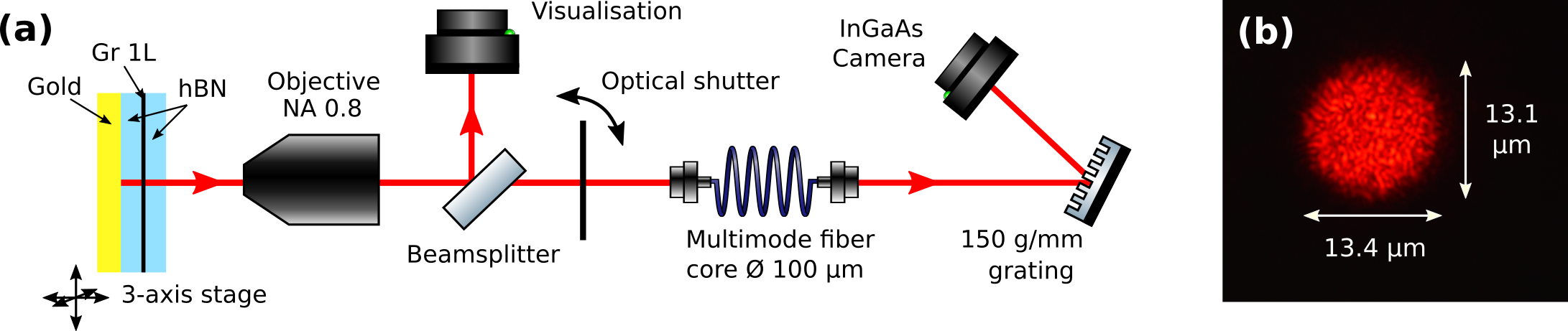}
\caption{a) Simplified experimental setup. b) Reciprocal image of the multi-mode fiber core on the sample focal plane.\label{fig_SI_Marin_1}}

\end{figure}

The blackbody emission is collected by a high numerical aperture objective, collimated via a 100~$\mu$m core-diameter multi-mode fiber, and sent to a NIR spectrometer providing a wavelength range of about 250~nm with a resolution of 1~nm (Fig.\ref{fig_SI_Marin_1}a). The \cv{measured}{} signal is integrated \cv{over a time-lapse of}{for} 100 seconds \cv{, significantly increasing the signal at a temperature of about 500~K}{}.

As the incandescent emission is collimated by the multi-mode fiber, only an effective surface of the sample $S_{\mathrm{collection}}$ contributes to the detected signal. This collection region is the reciprocal image of the multi-mode fiber core created by the optical system and can be measured by back-injecting some light from the fiber to the sample focal plane (all optical elements are apochromatic), as illustrated in Fig.~\ref{fig_SI_Marin_1}b. 

The measured spectral flux (in units of $\rm{W.m^{-1}}$) can thus be described by the following formula:
$S(\lambda, T)=R_{\mathrm{col}}(\lambda) S_{\text {emitter }} \varepsilon(\lambda) B(\lambda, T)$ with $B(\lambda, T)=\frac{4 \pi h c^2}{\lambda^5} \frac{1}{\exp \left(h c / \lambda k_B T\right)-1}$ the spectral blackbody radiance integrated over a half-space ($\lambda$ is the emission wavelength, $T$ is the temperature, $h$ is Planck's constant, $c$ is the celerity of light, and $k_B$ is Boltzmann's constant), $R_{\mathrm{col}}(\lambda)$ is the calibrated collection efficiency \eb{}{of the complete detection chain}, $S_{\text {emitter }} < S_{\text {collection}} = 138\,\rm{\mu m^2}$ is the surface from which radiation is emitted and $\varepsilon(\lambda)$ the emissivity. 

\begin{figure}
\centering
\includegraphics[width=.8\linewidth]{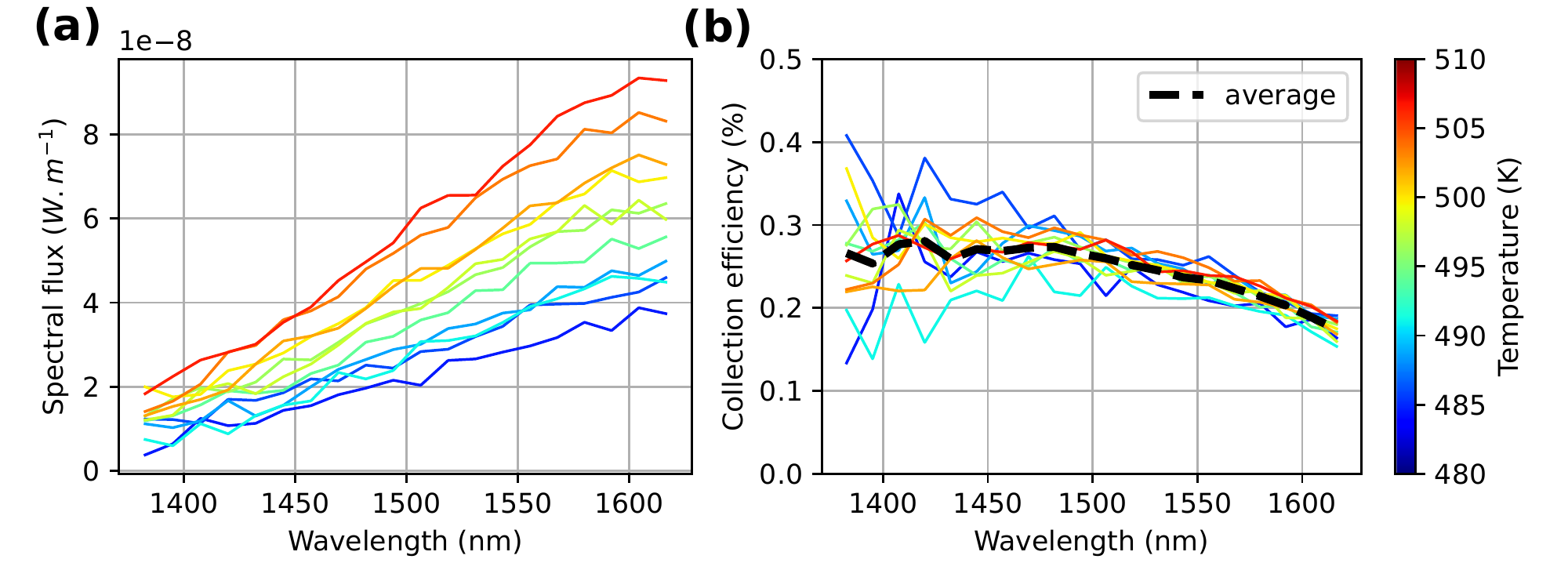}
\caption{(a) Measured spectral flux radiated from the tungsten calibration plate. (b) Obtained collection efficiency curves over a temperature range from 480 K to 510 K. The average of these curves is used for the calibration.\label{fig_SI_Marin_2}}

\end{figure}

The spectral flux is calibrated using a heated tungsten plate, of known emissivity [\onlinecite{pon1984spectral}], whose temperature is measured with a Pt100 temperature probe, and emitter surface is equal to the full collection surface $S_{\mathrm{emitter}}=S_{\mathrm{collection}}$. This measurement is performed at different temperature setpoints (Fig.~\ref{fig_SI_Marin_2}a), giving overlapping collection efficiency profiles (Fig.~\ref{fig_SI_Marin_2}b). The calibrated collection efficiency $R_{\mathrm{col}}(\lambda)$ is found by taking the average of these profiles.

\begin{figure}
\centering
\includegraphics[width=\linewidth]{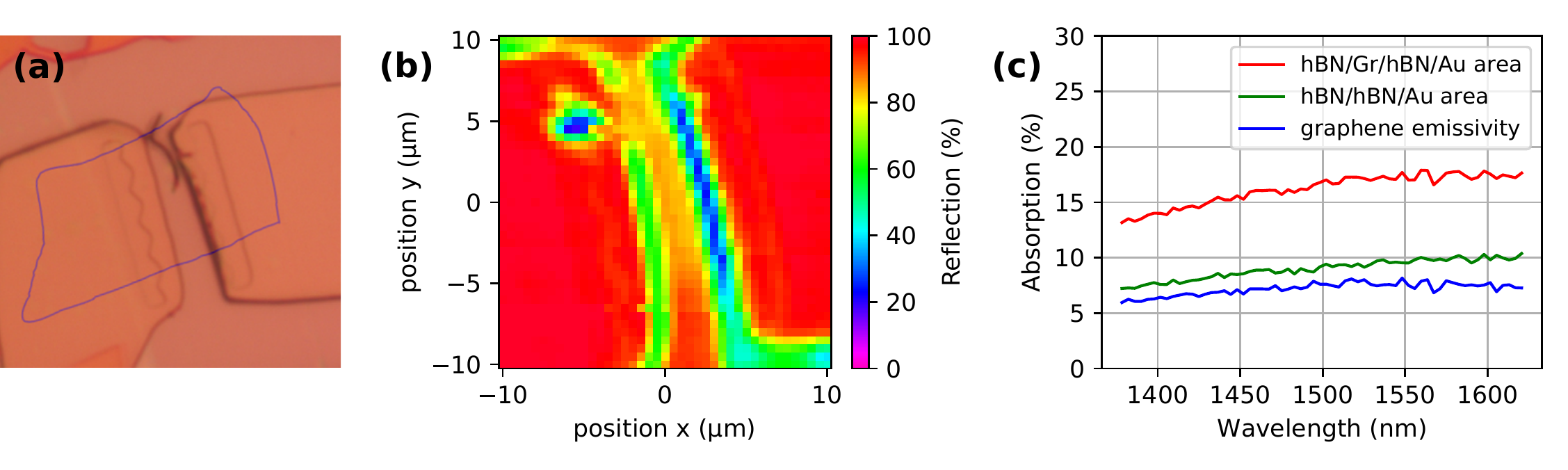}
\caption{(a) Image of the F160 transistor, with the graphene flake area highlighted in blue. (b) Absolute reflection of the device at 1500 nm. The gold contacts appear in red with a $98\%$ reflection while the graphene area in between the contacts appears orange with a reflection of around $85\%$. (c) Normalized absorption of the hBN/graphene/hBN/Au and hBN/hBN/Au stacks with their difference giving that of the transistor's emissivity (blue curve)
. 
\label{fig_SI_Marin_3}}

\end{figure}

\subsection{Emissivity measurement}
Under bias, the gold backgate temperature is significantly lower than that of graphene's electron temperature so that only the latter has a significant contribution to the blackbody emission. However, the emissivity of the graphene sheet cannot be directly computed from a simple thin film transfer matrix method as the radiation is collected with a high numerical aperture microscope objective from a stack that is several hundreds of nanometers thick. Therefore, instead of relying on an idealized simulation, we rather perform a spectral absorption measurement of the stacked structure by using the same microscope objective and by focusing a super-continuum laser on the sample focal plane.
The power reflected off the surface is normalized using the signal measured on the transistor's gold contacts, which has a tabulated reflection [\onlinecite{johnson1972optical}] (Fig.~\ref{fig_SI_Marin_3}a and \ref{fig_SI_Marin_3}b). For scatterer-free interfaces, the absorbed fraction of light $A$ is given by $A = 1 - R$ where $R$ is the measured reflection.
We then measure the absorption of the hBN/graphene/hBN/Au heterostructure as well as that of the hBN/hBN/Au stack where the graphene sheet is absent. It should be noted here that both graphene and gold contribute to the absorption of such structures, due to their finite extinction coefficients. Since the graphene sheet is optically thin, its spectral emissivity can be approximated by taking the difference between the absorption of the hBN/graphene/hBN/Au and hBN/hBN/Au stacks (Fig.~\ref{fig_SI_Marin_3}c).

\subsection{Temperature fitting}
The spectral blackbody irradiance emitted from the transistor's graphene channel over a half-space (in units of $\rm{W.m^{-3}}$) can thus be computed by normalizing the measured spectral flux by the collection efficiency $R_{\mathrm{col}}(\lambda)$, and the transistor's surface $S_{\mathrm{emitter}}$. This irradiance is then fitted with $\varepsilon(\lambda) B(\lambda,T)$, the measured emissivity of the embedded graphene sheet times the spectral blackbody irradiance, with the temperature $T$ as the only free parameter (Fig. 1C of the main text).

\begin{figure}[ht!]
\centering
\includegraphics[width=\linewidth]{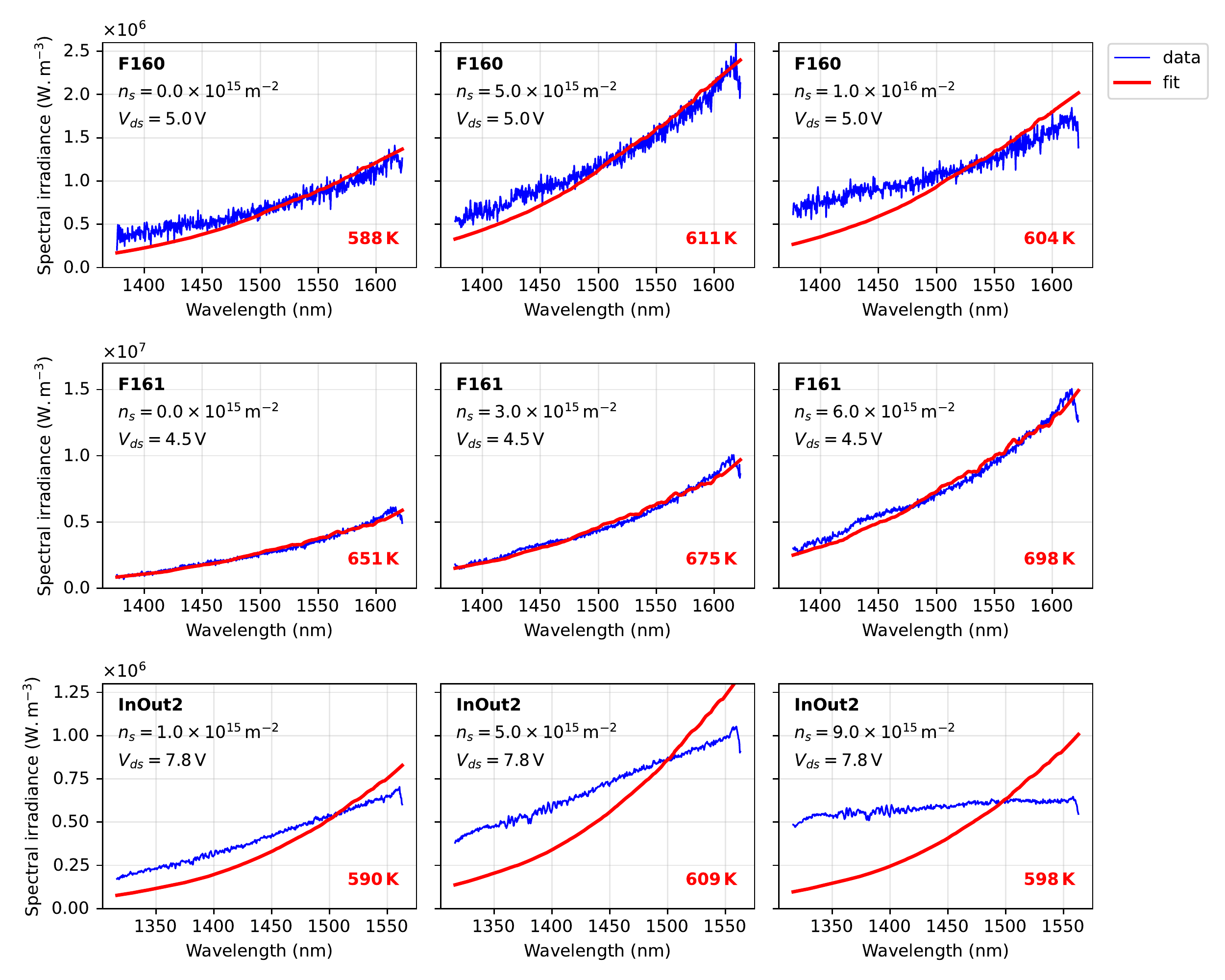}
\caption{Spectral irradiance of F160, F161, and InOut2 (top to bottom), for increasing doping (left to right) in the NIR over the spectral range $1.3\, -\, 1.6\, \rm{\mu m}$. Experimental data are fitted with the greybody model taking into account the measured device emissivity as described in the text. This one-parameter fit allows determining the electronic temperature for most doping values with F160 and F161. The good agreement of the fit with experimental data supports the assumption of a quasi-uniform channel temperature. In contrast, the short wavelengths are over-represented for InOut2 at moderate and large bias, indicating \eb{the presence of subwavelength hot spots in the channel}{a nonuniform temperature in the channel. In such cases,  NIR temperature data are systematically discarded}. 
}
 \label{fig_SI_Marin_4}
\end{figure}

The fit performs well in general on the F161 and F160 transistors (see Figs.~\ref{fig_SI_Marin_4}). However, at large doping, the fit deviates significantly from F160's measured irradiance  (Fig.~\ref{fig_SI_Marin_4}). This effect could be attributed to temperature inhomogeneity of the channel, \emph{i.e.}, a modification of the temperature profile along the channel with the doping, combined with a damped emissivity close to the contact area, which can modify the slope of the measured spectral irradiance. This temperature inhomogeneity is exacerbated in the long transistor InOut2 for which fits perform correctly at moderate bias or low doping and fail at large bias and large doping so that NIR blackbody thermometry fails in this case.
Moreover, the contact edges are responsible for significant light scattering, where reflection can reach $20 \, \%$ (see Fig.~\ref{fig_SI_Marin_3}b).
 Nonetheless, the exponential dependency of radiated power on absolute temperature \eb{limits the fit's deviation to a temperature uncertainty}{strongly constraints the determination of temperature and limits its uncertainty to} \eb{of}{} only 5~K over most of the explored doping and bias range. The worst cases,  at large doping and large bias, are $-5/+25$~K for F160 and $-20/+50$~K for InOut2, both being represented on the right panels.


\section{Stokes-anti-Stokes Raman thermometry}

We introduce the experimental methods for the determination of the temperature of optical phonons using Stokes-anti-Stokes (SAS) Raman thermometry. This technique enables a precise and unequivocal determination of the temperature of optical phonons, in contrast to techniques that only rely on the shift of the Stokes peaks, which is also sensitive to doping and strain. 

\begin{figure}[h!]
\centering{}\includegraphics[width=13cm]{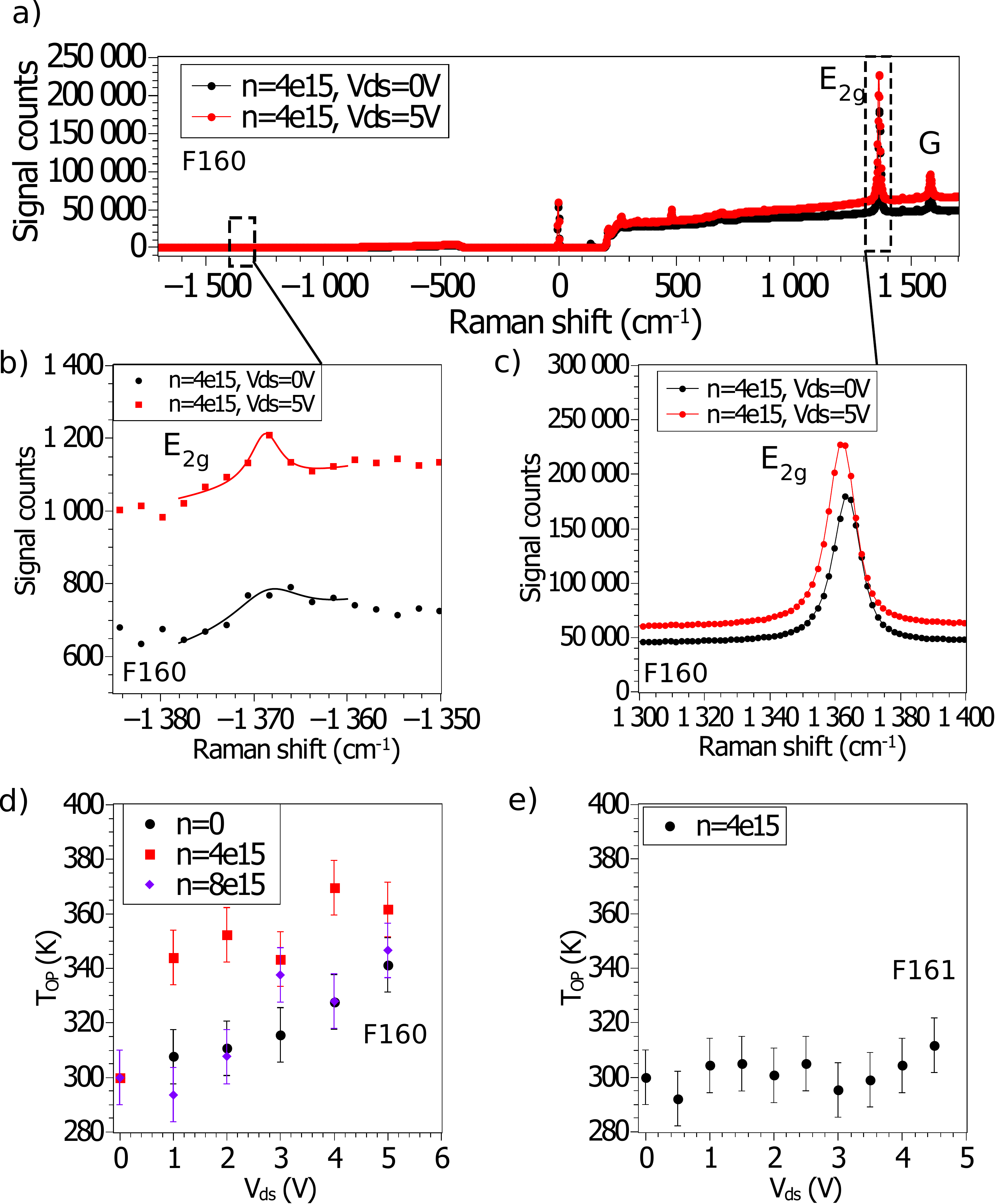} \caption{ Stokes-anti-Stokes Raman thermometry of high-mobility graphene transistors. a) Typical Raman spectra over the whole measurement range for zero bias (black data points) and high bias (red data points). The graphene G peak ($1580 \mathrm{cm}^{-1}$) and the hBN $E_{2g}$ peak ($1365 \mathrm{cm}^{-1}$) are clearly visible on the Stokes side. Panels b) and c) show a zoom on the antiStokes and Stokes hBN $E_{2g}$ peak, respectively, along with Lorentzian fits.  d) Associated temperature of the optical phonons of hBN for F160 as a function of bias for three values of doping $n=[0,0.4,0.8] \times 10^{16} \, \mathrm{m}^{-2}$. e) Temperature of the optical phonons of hBN versus bias for F161 at $n=0.4 \times 10^{16} \, \mathrm{m}^{-2}$. }
\label{SAS}
\end{figure}

We use a \emph{Renishaw Raman} spectrometer with a notch filter that allows simultaneous measurement of Stokes (S) and anti-stokes (AS) peaks. The transistor is excited with a $\lambda = 532 \, \mathrm{nm}$ laser ($P=5 \, \mathrm{mW}$). Typical spectra are shown in Fig.~\ref{SAS}-a, with a zoom on the hBN $E_{2g}$ peak on the anti-Stokes and Stokes sides in panels b and c, respectively.

We rely on the intensity ratio between the S and AS peaks at a temperature $T$ given by $\frac{I_{AS}}{I_{S}}=e^{-\frac{\hbar \Omega}{k_{B}T}}$, where $\hbar \Omega$ is the energy of the optical phonon mode. For the hBN optical phonon mode $E_{2g}$ at $1365~\mathrm{cm}^{-1} = 169~ \mathrm{meV}$ , this yields a ratio of $\sim 10^{-3}$ at room temperature. Yet the AS peaks can be seen even at zero bias voltage by using a long exposure time, as can be seen in Fig. \ref{SAS}b (black data points). This measurement is used as an absolute calibration for SAS thermometry. 

By comparing the AS/S ratios at zero bias and non-zero bias, we deduce the temperature of the optical phonons of hBN as a function of doping and bias voltage. Stokes-anti-Stokes measurements were performed on two electroluminescent graphene transistors (F160 and F161), and the associated temperature is presented in Figs. \ref{SAS}d) and e). The temperature of hBN's optical phonons shows a slight increase as a function of electrical bias. As such, it is necessary to verify that the emitted MIR signal presented in the main text is not due to incandescence from the hBN environment. However, when placing the heterostructure on a hot plate at a temperature that corresponds to the one measured with SAS thermometry, we observe that the incandescent emission from the heterostructure at $1348 \, \mathrm{cm}^{-1}$ is negligible (see main text). This excludes the hot hBN as the MIR emitter.

\section{Devices}
\textit{Electronic transport characterization}. The electronic transport characterization of the high-mobility graphene transistors is performed prior to optical measurements at room temperature in a probe station adapted to radio frequency measurements up to 67~GHz. The DC measurements are performed using a \emph{Keithley} 2612 voltage source to apply gate and bias voltages. 
When applying a constant gate voltage and increasing the bias voltage, the average doping in the graphene channel decreases due to drain doping. In our measurements, we correct for this effect by biasing the graphene transistors along constant density lines ($V_{gs}- \alpha V_{ds} = \mathrm{Const.}$, where $\alpha\sim 0.4$ is determined from the shift of the charge neutrality point with respect to bias [\onlinecite{Yang2018nnano}]). 

\textit{Electronic properties of the devices}. We describe here the different electronic transport regimes of the 3 high-mobility transistors detailed in this study: F160, F161, and InOut2. Their geometrical properties and mobility are listed in table \ref{table:devices}.

\begin{table}[t]
    \centering
    \begin{tabular}{cccccc}
    \hline
    \hline
       Device & Dimensions  & Clear aperture & hBN thickness & Electronic mobility & Contact resistance \\
      & $L \times W$ ($\rm{\mu m \times \mu m }$) & 
      $L \times W$ ($\rm{\mu m \times \mu m }$) &
      (top/bottom) ($\rm{nm / nm }$) & ($\textrm{m}^2.\textrm{V}^{-1}.\textrm{s}^{-1}$) & ($\Omega$) \\
        \hline
       F160 & $8.3 \times 12.5$ & $5.5 \times 12.5$ &$45/160$ & $13$ & $81$ \\
     F161 & $6\times 10$ &$5 \times 10$ & $54/152$ & $15$ &  $51$\\
     InOut2 & $20\times 8.5$ & $17 \times 8.5$ & $67/118$ & $9$ & $300$ \\
       \hline
       \hline
       \end{tabular}
    \caption{Dimensions and transport properties of the three high-mobility graphene transistors under study.}
    \label{table:devices}
\end{table}

\begin{figure}[h!]
\hspace*{-1cm}
\centering{}\includegraphics[width=17cm]{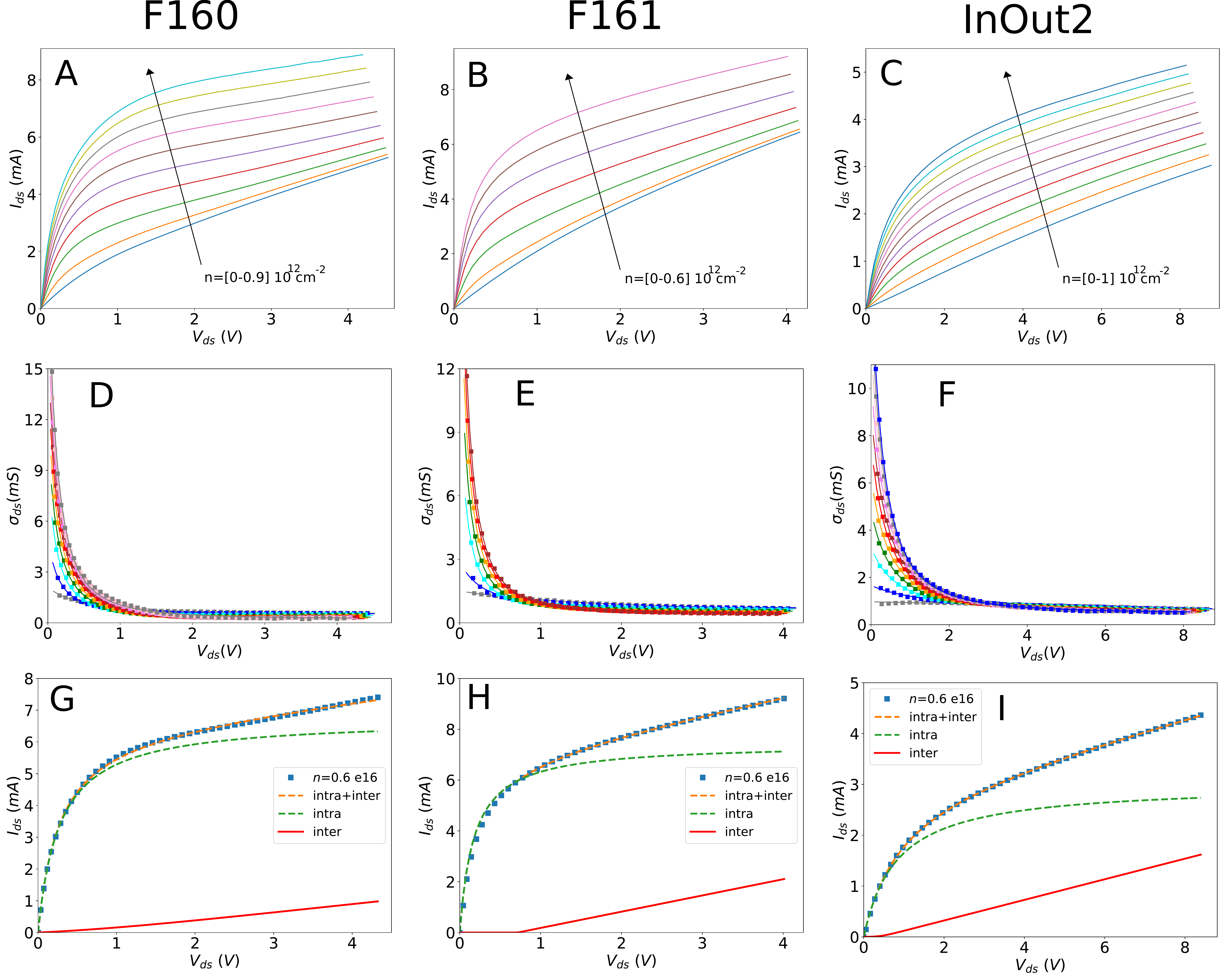} \caption{Transport properties of the electroluminescent graphene transistors : F160 (panels A,D and G), F161 (panels B,E and H), and InOut2 (panels C,F and I). For each device, the current-voltage curves are displayed for various dopings (panels A,B,C) between $n=0$ and $n=1~10^{12}\mathrm{cm^{-2}}$, as well as the differential conductivity $\sigma$ as function of bias in panels D,E,F (dots) along with the fits (solid lines) using Eq. (\ref{eq1}). In panels G,H,I  we perform a fit of the IV curves taking into account the existence of a Zener threshold, where $V_{Z}$ and $T_{e}$ are left as free parameters. The total current fit is plotted in orange, whereas the intraband current appears in green and underlines the quasi-linear increase of interband current at high bias (confirmed by the red solid line representing the interband contribution).}
\label{devices}
\end{figure}

The current-voltage curves at constant charge carrier density are plotted in Fig.~\ref{devices} for the three devices on a gold backgate (panels A,B,C). At low bias, the devices show linear ohmic behavior, with a strong increase of the intraband current with doping. Due to the high mobility, the current rapidly reaches significant values and saturates for $V_{ds} > V_{sat}$, where $V_{sat}$ is the saturation voltage. This saturation of intraband current can be consistently attributed to the interaction with optical phonons of the lower \emph{Reststrahlen} band of hBN ($\hbar \Omega_{I}= 90-100$~meV) [\onlinecite{Yang2018nnano},\onlinecite{Baudin2020adfm}].
This velocity saturation regime is followed at high bias by the onset of Zener interband conduction, characterized by a constant bias- and doping-independent differential conductivity $\sigma_{Z}\lesssim1\;\mathrm{mS}$, and the onset of energy relaxation and cooling by optical phonons of the upper \emph{Reststrahlen} band of hBN ($\hbar \Omega_{II}= 170-200 $~meV) [\onlinecite{Yang2018nnano,Baudin2020adfm}]. 
The differential conductivity can be well fitted using the standard dependence [\onlinecite{Yang2018nnano},\onlinecite{Meric2008}]: 
\begin{equation}
\label{eq1}
    \sigma(E)= \frac{ne\mu}{(1+E/E_{sat})^{2}}+ \sigma_{Z}
\end{equation}
where $\mu$ is the electronic mobility at zero bias, $n$ is the charge carrier density, and $E=V/L$ is the local electric field. Fits of the differential conductivity are presented in panels D,E,F of Figure \ref{devices}, showing good agreement with data. 

The standard dependence of Eq.(\ref{eq1}) neglects the existence of a doping-dependent threshold for the onset of interband Zener conduction due to Pauli blocking, which has been confirmed previously in noise temperature measurements [\onlinecite{Yang2018nnano}]. The fair agreement with the data upon neglecting the threshold is because the Zener threshold only involves minute corrections to conductance in high-mobility devices at low bias, where intraband conductance dominates. A better description relies on an optical conductivity-like formula for the interband conductivity as a function of bias $\sigma_{ds,inter}=\frac{\sigma_{Z}}{2} \bigl[ 1+ \text{tanh} \bigl( \frac{e(V_{ds}-V_{Z})}{2k_{B} T_{e}} \bigr) \bigr]$ where $V_{Z}$ is the threshold Zener voltage, and $T_{e}$ is the electronic temperature; $\sigma_{Z}$ is still extracted from the slope of the current-voltage curves at high bias. Combining this equation for interband conduction with the intraband part of Eq.(\ref{eq1}) yields the following formula for the total bias current  


$\left\{
\begin{array}{ll}
      I_{ds}=I_{intra}+I_{inter} \\
     I_{intra}=\frac{W}{L}ne\mu \frac{V_{sat}~V_{ds}}{V_{sat}+V_{ds}} \\
     I_{inter}=\frac{k_{B}T_{e}}{e} \sigma_{Z} \frac{W}{L} \biggl[ \frac{e(V_{ds}-V_{Z})}{2k_{B}T_{e}} +\text{ln}\bigl[ \text{cosh}\bigl(\frac{e(V_{ds}-V_{Z})}{2k_{B}T_{e}} \bigr) \bigr] +\frac{eV_{Z}}{2k_{B}T_{e}}-\text{ln}\bigl[ \text{cosh}\bigl(\frac{eV_{Z}}{2k_{B}T_{e}} \bigr) \bigr] \biggr]
\label{formule_complete}
\end{array}
\right.$

\vspace{0.5cm}

Using this formula, we perform fits on the current-voltage curves of the electroluminescent graphene transistors. Due to the fair agreement of the $\sigma$-fits with Eq. (\ref{eq1}), the values of $V_{sat}$, $\mu$ and $\sigma_{Z}$ are set, and only $V_{Z}$ and $T_{e}$ are left as adjustable parameters. The result is presented in panels G,H,I of Fig. \ref{devices} (orange lines) and shows excellent agreement with the data (blue dots), with values for $T_e  \simeq T_{e}(V_Z) \sim 600$~K (in this simple description, we consider a constant value of electronic temperature $T_e$, which can be seen as the electronic temperature close to the threshold voltage $V_Z$) and $V_{Z} \lesssim 0.5$~V that are compatible with optical measurements in the near- and mid-infrared. The intraband part of the current (green lines) shows a quasi-complete saturation for $V_{ds} \gg V_{sat}$, resulting in a linear increase of interband current at large bias (red solid line).  

\section{MIR absorption and emission spectra of high-mobility graphene transistors: the Local Kirchhoff Law}

\subsection{Experimental MIR emission spectra of graphene transistors}

In this section, we first report on the experimental observation of MIR radiative emission under a large bias and compare it to incandescent emission when the whole device is heated. We observe consistent emission at the quarter-wave resonance of the structures ($1348\, \mathrm{cm^{-1}}$).

\begin{figure}[h!]
\hspace*{-1cm}
\centering{}\includegraphics[scale=0.8]{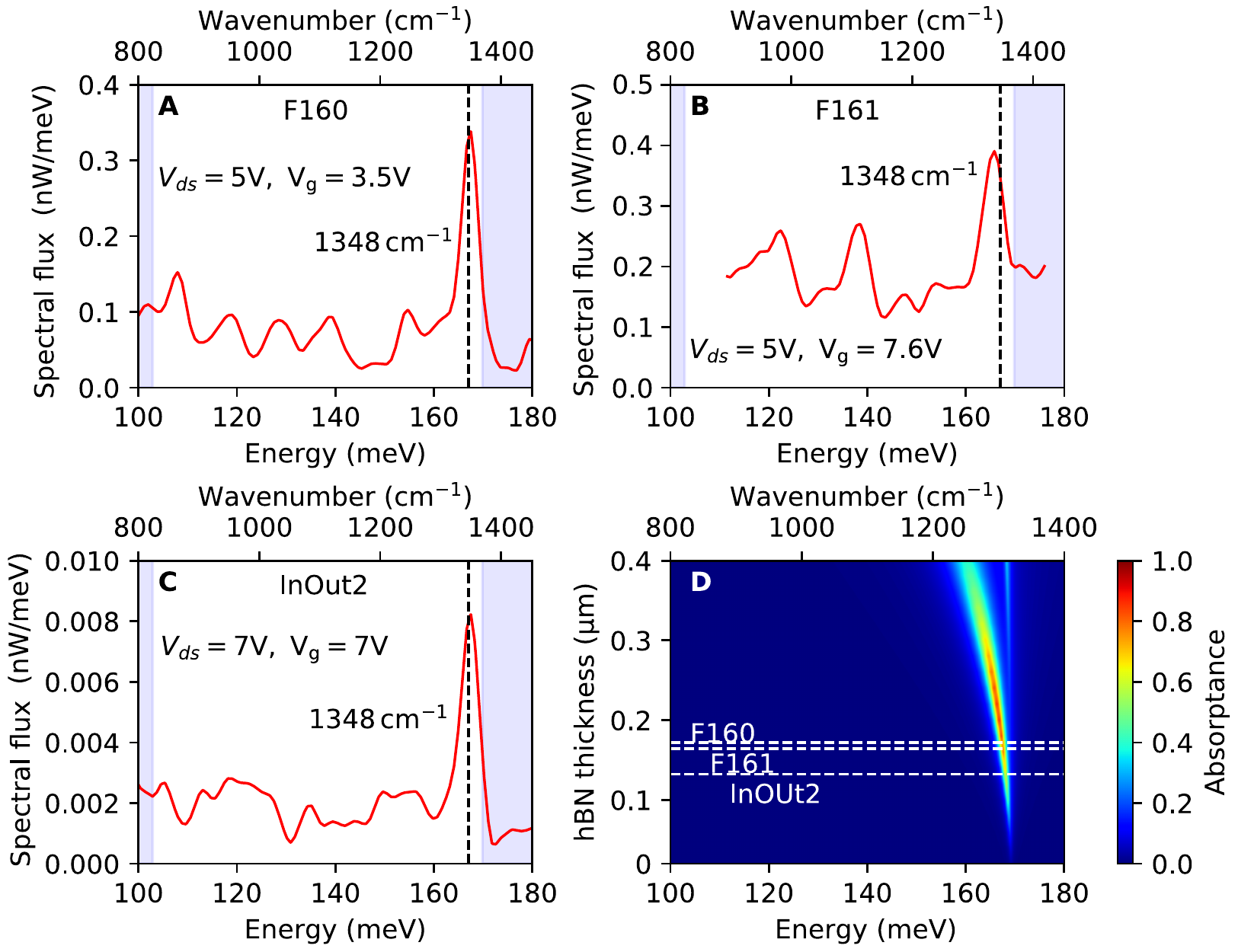}
\caption{Mid-infrared emission spectra of F160 (A), F161 (B), InOut2 (C), when submitted to a large bias. The reference $1348\,\rm{cm^{-1}}$ dashed line is represented for comparison. (D) Simulation of the emission peak energy function of the total hBN thickness. Dashed lines correspond to the hBN bottom thicknesses of the various HGFETs. }
\label{fig:EmissionPeaks}
\end{figure}

\begin{figure}[t!]
\hspace*{-1cm}
\centering{}\includegraphics[width=15cm]{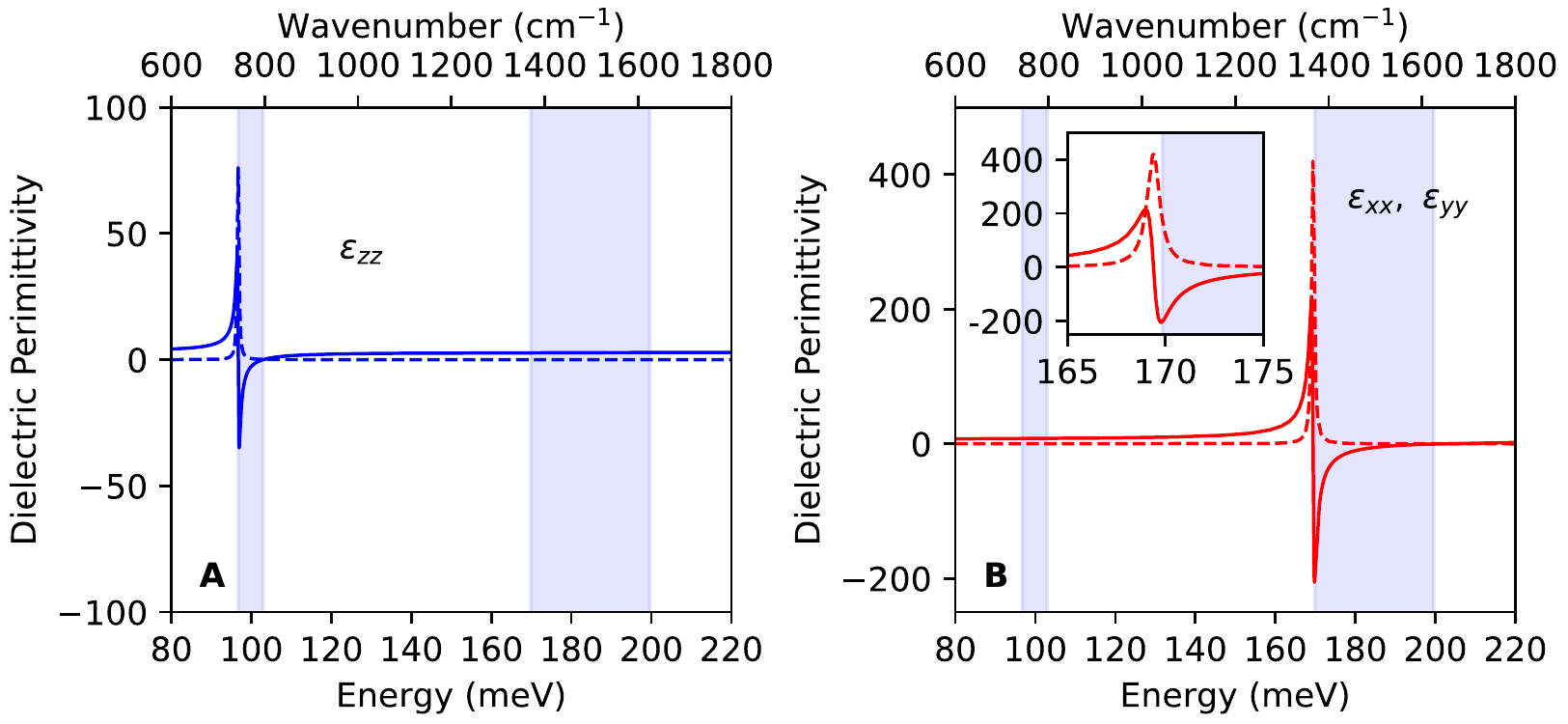}
\caption{(A) Out-of-plane and (B) in-plane dielectric permittivity of hBN using the model described in Eq.~(\ref{eq:temp_hBN}). Solid lines: real part, dashed lines: imaginary part. The shaded regions mark the material's two \emph{Reststrahlen} bands, in which the real part of the dielectric permittivity is negative. (Inset) Zoom on the in-plane OP resonance.}
\label{hBN_dielec}
\end{figure}

Figure~\ref{fig:EmissionPeaks} represents the measured emission spectra of the three devices reported in this study. The emission peak emerges from the baseband instrumental noise at $1348 \, {\rm cm^{-1}}(\equiv167 \, {\rm meV})$. 
Using S matrix calculations (see below), we compute the hBN total absorption (which coincides with emissivity) as a function of its thickness. The hBN layer in-plane dielectric response has a pronounced resonance peak at the in-plane transverse optical phonon frequency ($1370\, \rm{cm^{-1}}$, see Fig.~\ref{hBN_dielec}). 
The high refractive index near this frequency enables a quarter-wave resonance. 

We find that maximal hBN absorptance is reached for a total thickness of $\simeq 200\, \mathrm{nm}$ corresponding to that of our transistors. The quarter-wave resonance redshifts with the thickness of the hBN layer as a consequence of its dispersion. Interestingly, according to this numerical calculation, for large hBN thicknesses, a second resonance corresponding to a $3\lambda/4$ resonance appears.

The transverse electric field distribution in the case of F160 dimensions is presented in the inset of Fig. 1C. Graphene absorption (and therefore emissivity) is maximized by placing it near the maximal electric field which is near the air-hBN interface. This is why we kept the top hBN thickness as thin as possible in all the devices considered here while securing encapsulation to ensure operation in ambient conditions.

\begin{figure}[h!]
\hspace*{-1cm}
\centering{}\includegraphics[scale=0.8]{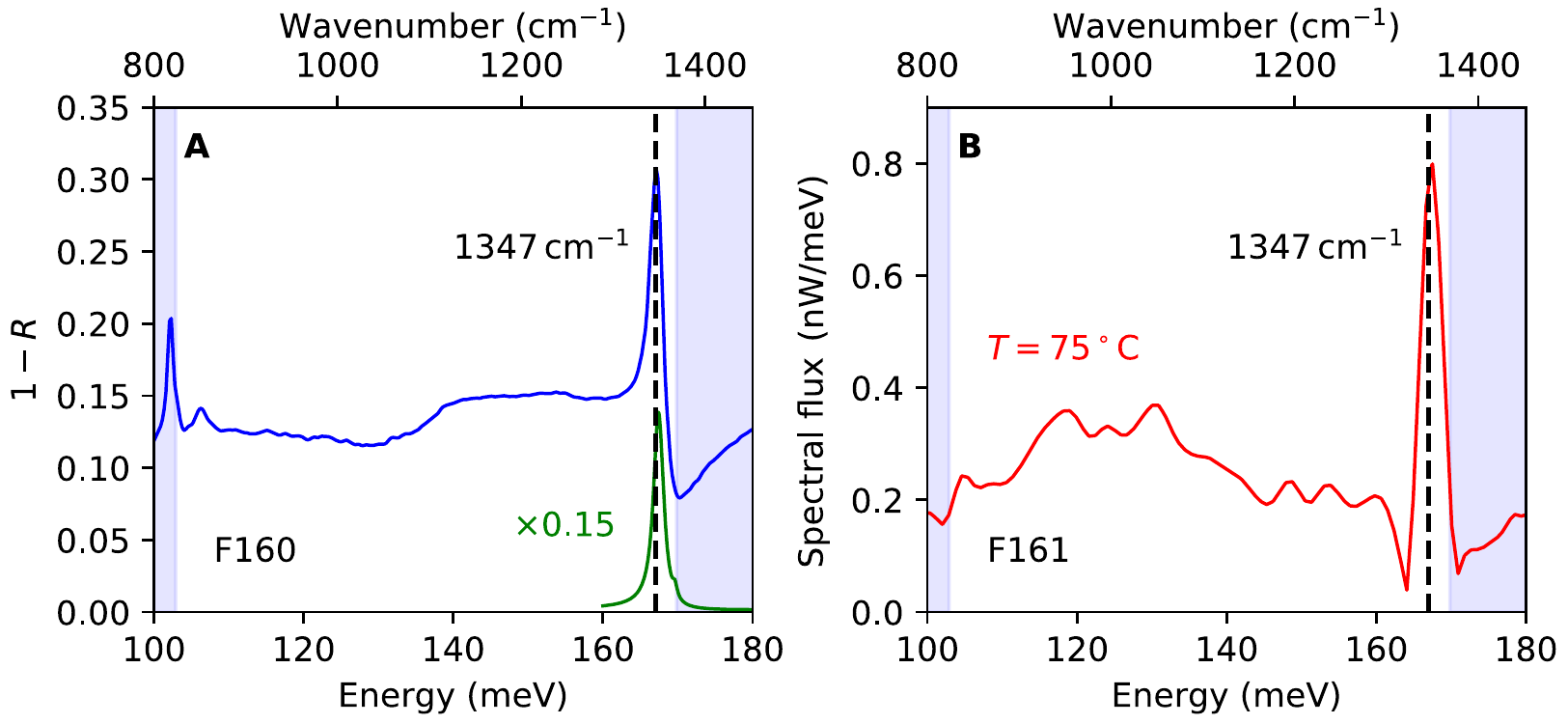}
\caption{(A) Experimental absorptivity (blue) spectrum of F160, and computed emissivity (green) rescaled to match the peak amplitude. 
(B) Incandescent emission spectrum from F161 heated to 75°C, measured via IR-SMS. }
\label{fig:hBNPeakMeas}
\end{figure}

We independently verify the existence of the \eb{}{resonance} peak computed in Fig.~\ref{fig:EmissionPeaks}D in two ways, which we illustrate in what follows. (i) The MIR absorptivity $A=1-R$ (see subsection~\ref{MIR refl}) of F160, in which the absorption peak \eb{of hBN}{} is clearly discernible (see Fig.~\ref{fig:hBNPeakMeas}A). 
(ii) We can also observe this \eb{hBN}{} quarter-wave mode through its incandescent emission obtained by heating the F161 device to a high temperature, as illustrated in Fig.~\ref{fig:hBNPeakMeas}B.
\eb{}{Note that, since hBN and graphene absorption/emission peaks coincide (see Fig.~\ref{F160_decomp}), the quantitative comparison of absorption and incandescent emission is not straightforward as, under electrical bias, only the graphene layer emits, whereas in the incandescent case by external heating, emission is dominated by the hBN layers, their out-coupling strength being 47 times larger than the one of graphene (see below). The solution to this issue is provided by the local Kirchhoff law.}

\subsection{The Local Kirchhoff Law}
\subsubsection{Presentation}

We model the out-of-equilibrium electroluminescent emission of a graphene field effect transistor using the local Kirchhoff law (LKL)[\onlinecite{greffetLightEmissionNonequilibrium2018}] which provides an expression for the spectral radiance within a solid angle $d\Omega = \sin{\theta} d\theta d\phi$ as follows 

\begin{equation}
\frac{ \rm{d}^2 P_{e}}{\rm{d}\omega\ \rm{d}\Omega}=\frac{1}{2}\sum_{l} \int\alpha^{(l)}\left(-\mathbf{u}_{r},\mathbf{r}^{\prime},\omega\right)I_{b}\left[n_{e}\left(\omega,\mathbf{r}^{\prime}\right),\omega\right]\rm{d}^{3}\mathbf{r}^{\prime}\label{eq:GKL}
\end{equation}

\noindent
where $\alpha_{\mathrm{ib}}^{(l)}\left(-\mathbf{u},\mathbf{r}^{\prime},\omega\right)$
is the polarized absorption cross-section density due to interband transitions
(direction ${\bf u}$, frequency $\omega$, and polarization $l$),
and $I_{b}\left[n_{e}\left(\omega,\mathbf{r}^{\prime}\right),\omega\right]$
is the local body radiance, which for interband transitions is given by
$I_{b}\left[n_{e}\left(\omega,\mathbf{r}^{\prime}\right),\omega\right]=\frac{\omega^{2}}{4\pi^{3}c^{2}}\hbar\omega~n_{e}\left(\omega/2,\mathbf{r}^{\prime}\right)n_{h}\left(-\omega/2,\mathbf{r}^{\prime}\right)$ and is just a bimolecular recombination rate of electron-hole pairs,
with hole probability $n_{h}=1-n_{e}.$ At thermal equilibrium, this expression reduces to

\begin{equation}
    I_{b}\left[n_{e}\left(\omega,\mathbf{r}^{\prime}\right),\omega\right]=\frac{\omega^{2}}{4\pi^{3}c^{2}}\frac{\hbar\omega}{\exp\left(\frac{\hbar\omega}{k_{B}T}\right)-1}. \label{therm eq}
\end{equation}
It is convenient to cast the non-equilibrium expression in the form of Eq.~(\ref{therm eq}) by incorporating the photon chemical potential $\Delta \mu$ (which is possibly frequency dependent) as follows 

\begin{equation}
I_{b}\left[n_{e}\left(\omega,\mathbf{r}^{\prime}\right),\omega\right]=\frac{\omega^{2}}{4\pi^{3}c^{2}}\frac{\hbar\omega}{\exp\left(\frac{\hbar\omega-\Delta \mu}{k_{B}T}\right)-1}.
\end{equation}

The LKL assumes that light emission occurs via a single direct optical transition, which excludes multiple competing direct interband transitions and intraband transitions. This assumption is well satisfied in the case of graphene over the spectral range of interest as we shall see in the next subsection. 

\subsubsection{Absorption of complex layered structures}

\eb{}{
To compute the absorptance of the layered heterostructure, we rely on the Reticolofilm-stack free software. The Reticolo software computes the reflection and transmission of arbitrary stacks of anisotropic thin films using a product of S matrices. The contribution to absorptance of each layer is deduced from the local balance of the Poynting vector. The computation is vectorized over energy and incident angle, and the result is integrated over the observation angle of the Cassegrain objective (see subsection.~\ref{sec IRSMS}).
}

\noindent
When an HGFET is submitted to a bias, only the graphene layer emits because of its much higher temperature. However, when measuring the spectral absorption of the heterostructure, both graphene and hBN absorb. This distinction is correctly captured by the local Kirchhoff law, whereas the macroscopic Kirchhoff law fails to describe such a situation. 

\begin{figure}[t!]
\hspace*{-1cm}
\centering{}\includegraphics[width=13cm]{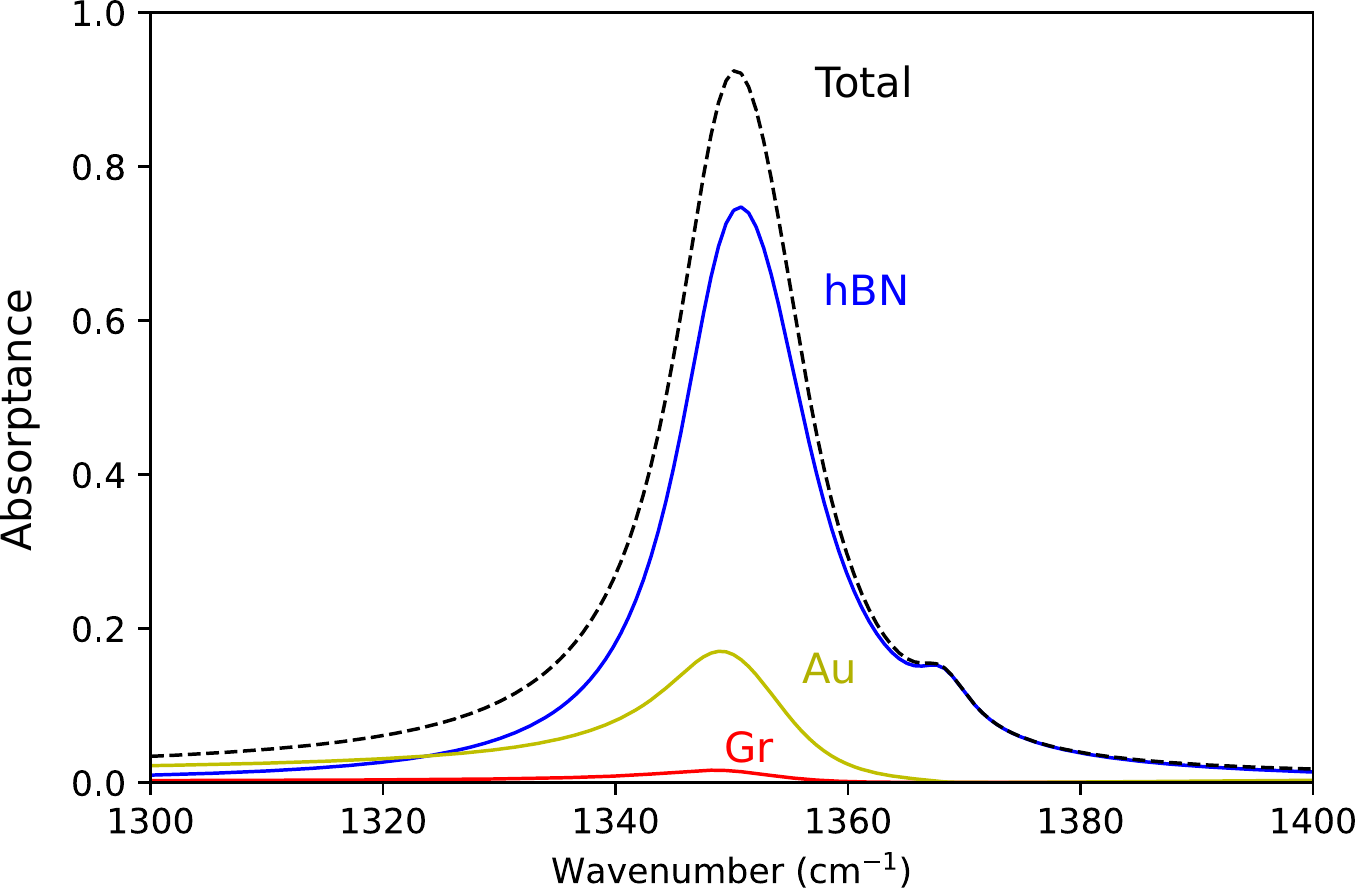}
\caption{Simulated F160 emittance spectrum integrated over the collection angles of the MIR microscope objective. Total emittance is the dashed black line and decomposes over the hBN (blue) the Au backgate (yellow), and the graphene contributions (red).}
\label{F160_decomp}
\end{figure}

Figure~\ref{F160_decomp} shows the decomposition of surface emittance (integrated over the microscope objective's angular collection interval) over the layers composing the hBN-encapsulated graphene heterostructure. If the device is uniformly heated, this decomposition provides the relative weight of each layer to the total emission (dashed line), in which the emission of hBN layers is predominant (blue). If, otherwise, only the graphene layer is emitting, the corresponding emittance is only a few percent (red curve). The average emittance of graphene over the entire emission bandwidth compared to the total heterostructure is found to be only 2.2\%. 
\eb{}{As an example, the Inset of Fig.~2A shows the spectral irradiance of F160 under external heating at $75^\circ\,\rm{C}$. In this case hBN dominates the total emission completely. If hBN layers were kept at room temperature, the same level of signal at resonance would be reached for a graphene temperature larger than $T_e \gtrsim 560^\circ\,\rm{C}$.}

\subsubsection{Modulation of absorptivity: thermochroism and electrochroism}

\begin{figure}[th!]
\hspace*{-1cm}
\centering{}\includegraphics[width=15cm]{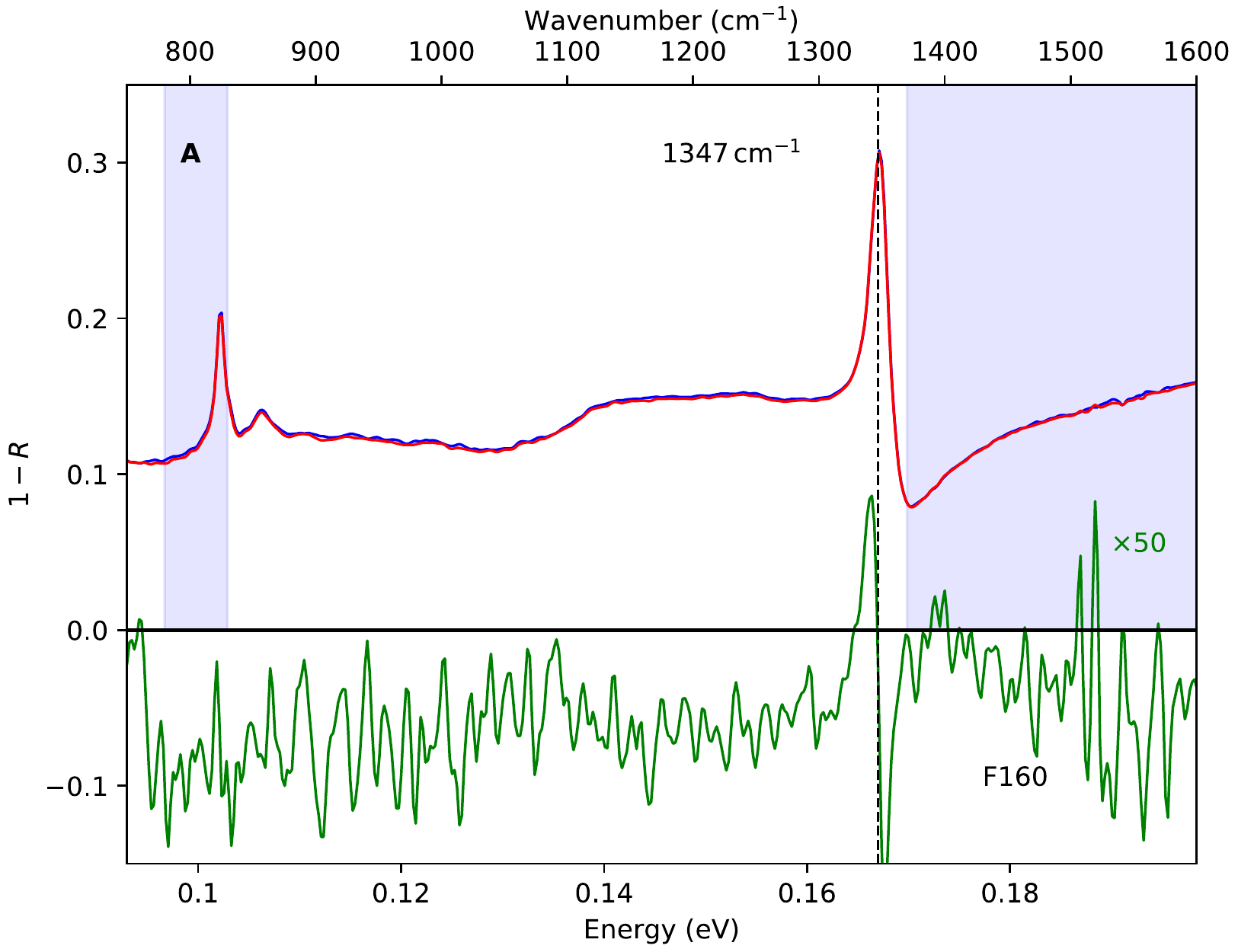}
\caption{Reflectance measurement of F160 at null bias (blue), and maximum bias (5V, red) to investigate electro- or thermochroism. The result is represented as the non-reflected fraction $(1-R)$ which coincides with absorptance is the absence of scattering. The difference between the biased case and the unbiased one is represented on the bottom green curve.}
\label{F160_electrochroism}
\end{figure}

For common blackbody or electroluminescent emitters, the absorption cross-section is constant with temperature or electric field. \eb{}{However, at low photon energies such as the energy of MIR photons, variations of temperature and electric field may induce a non-negligible effect on the absorptivity of materials. Thus, }using a MIR reflectance experiment, we verified that electro- or thermochroism are absent in HGFETs\eb{}{. We observe only} a tiny relative change in absolute absorptance of $+0.2\%$ (see Fig.~\ref{F160_electrochroism}) with increasing bias, which \eb{results}{leads to} a temperature increase.

\begin{figure}[t!]
\hspace*{-1cm}
\centering{}\includegraphics[width=15cm]{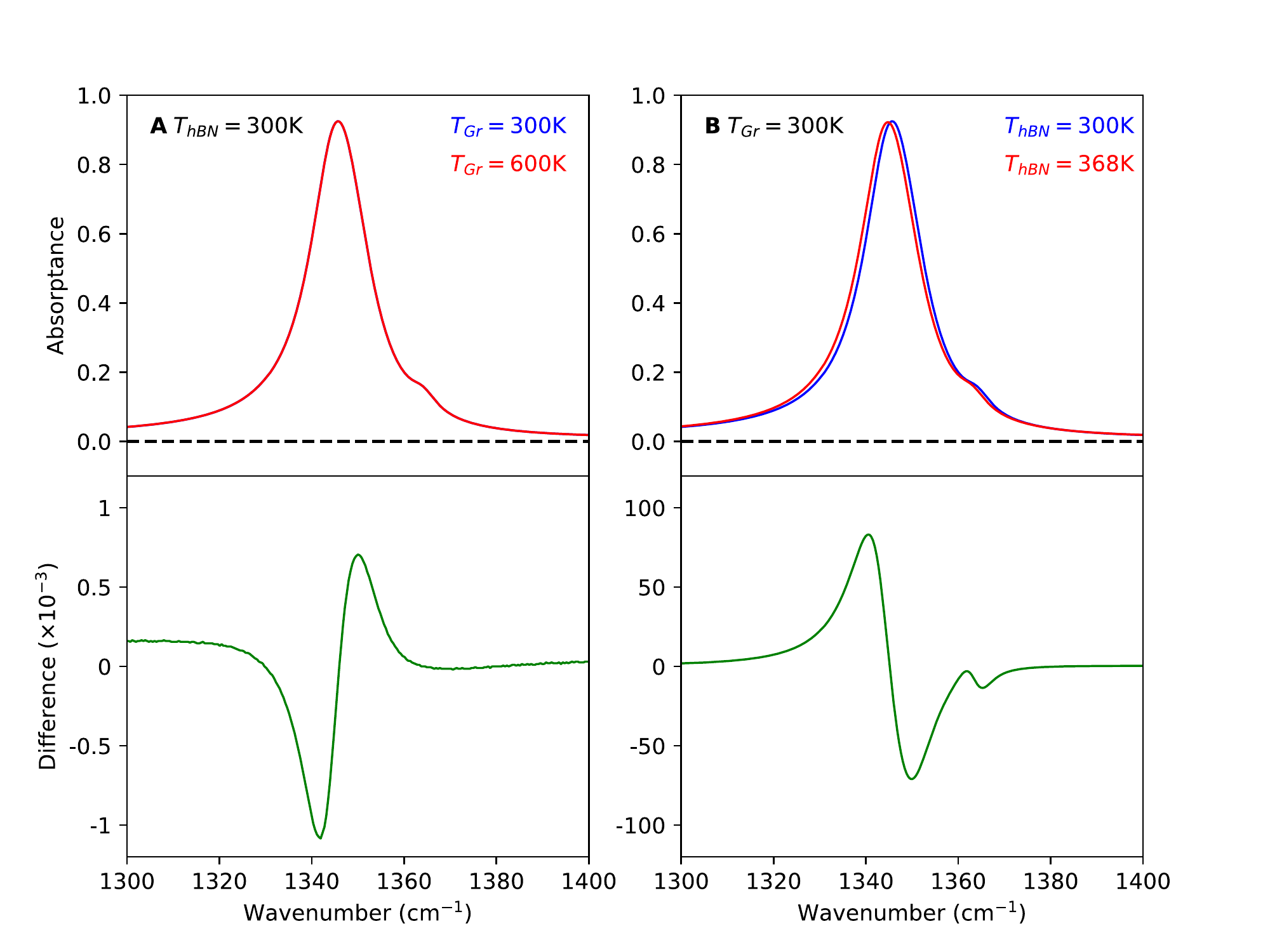}
\caption{
Computation of the influence of graphene temperature (a) and hBN temperature (b) on the absorptance peak. a) For graphene, hBN temperature is kept at 300K, while graphene temperature is increased from 300K (blue) to 600K (red) corresponding to the electronic temperature measured experimentally by NIR blackbody thermometry. The difference is represented below and consists mainly of a small blueshift of the resonance, and a relative change in absorptance of $-0.08\%$ b) For hBN, graphene temperature is kept at 300K, while hBN temperature is increased from 300K (blue) to 368K (red) corresponding to the hBN OP temperature measured experimentally by Stokes-antiStokes Raman thermometry. The difference is represented below and consists mainly of a redshift of the resonance and a relative change of absorptance of  $1.4\%$.}
\label{F160_thermochroism}
\end{figure}

If this change \eb{results}{were to originate} from the graphene layer, the corresponding maximal relative absorptance change would be $+0.2\% / 2.2 \% = +9\%$. This can be quantitatively ruled out by comparing the computed emittance of graphene at room temperature and at 600~K, which corresponds to the characteristic temperature of graphene's electrons under large bias, see figure \ref{F160_thermochroism}A. In the computations, the temperature of graphene's electrons is taken into account in the optical conductivity of graphene [\onlinecite{falkovsky2008optical, messina2013graphene}] $\sigma = \sigma_{\rm{intra}}+\sigma_{\rm{inter}}$, which is given by

\begin{equation}
\begin{array}{l}
\sigma_{\rm{intra}}(\omega)=\frac{i}{\omega+\frac{i}{\tau}} \frac{2 e^2 k_B T}{\pi \hbar^2} \log \left(2 \cosh \frac{\mu}{2 k_B T}\right) . \\
\sigma_{\rm{inter}}(\omega)=\frac{e^2}{4 \hbar}\left[G\left(\frac{\hbar \omega }{2}\right)+i \frac{4 \hbar \omega}{\pi} \int_0^{+\infty} \frac{G(\xi)-G\left(\frac{\hbar \omega}{2}\right)}{(\hbar \omega)^2-4 \xi^2} \rm{d} \xi\right],
\end{array}
\label{eq:sigma_intra_inter}
\end{equation}

where $G(x)=\sinh \left(\frac{x}{k_B T}\right) /\left[\cosh \left(\frac{\mu}{k_B T}\right)+\cosh \left(\frac{x}{k_B T}\right)\right]$. The conductivity depends on the chemical potential $\mu$ and on the relaxation time $\tau = 1.5\, 10^{-13}\, \rm{s}$. We observe a minor decrease in absorption, reaching $-0.1\%$ at the maximum.

To summarize, the relative change in absorptance cannot be attributed to graphene which naturally becomes \eb{}{more} transparent in the MIR when temperature increases, so that its absorptance should \textit{decrease} with temperature (see Fig.~\ref{F160_thermochroism}A). 

Therefore, we turn to hBN as the origin of this tiny thermally induced absorptance change. We perform the same analysis using the experimentally measured temperature for hBN (368~K), and a temperature-dependent dielectric model of hBN:

\begin{equation}
\epsilon_{ii} = 
\epsilon_{ii}^{\infty}\, 
\left(
\frac{\omega_{LO,i}^{2}-\omega^{2}-i\omega\Gamma_{LO,i}}{\omega_{TO,i}^{2}-\omega^{2}-i\omega\Gamma_{TO,i}}
\right)
\label{eq:temp_hBN}
\end{equation}

where $i\in\{ x,y,z \}$, for the $x$ component:  $\epsilon_{zz}^{\infty}=2.95\ ,\omega_{TO,z}=780\,\rm{cm^{-1}},\ \omega_{LO,z}=830\,\rm{cm^{-1}},\ \ensuremath{\Gamma_{LO,z}}=\ensuremath{\Gamma_{TO,z}}=4$, and for the $x$ and $y$ components: 
\begin{equation}
\begin{cases}
\text{\ensuremath{\epsilon_{xx}^{\infty}}=} & 4.87\\
\text{\ensuremath{\omega_{TO,x}}=} & 
1365.98\,\rm{cm^{-1}}\left(1 - 9.506\ 10^{-6} \times (T-300\,\rm{K}) - 2.26 \ 10^{-8} \times (T-300\,\rm{K})^2\right)\\
\omega_{LO,x}= & 
1616.14\,\rm{cm^{-1}}\left(1 - 10.58\ 10^{-6} \times (T-300\,\rm{K}) - 1.7044 \ 10^{-8} \times (T-300\,\rm{K})^2\right)\\
\Gamma_{TO,x}= & 
6.4521\,\rm{cm^{-1}}\left(1 + 7.085\ 10^{-4} \times (T-300\,\rm{K}) + 1.2595 \ 10^{-6} \times (T-300\,\rm{K})^2\right)\\
\Gamma_{LO,x}= & 
4.1226\,\rm{cm^{-1}}\left(1 + 10.24\ 10^{-4} \times (T-300\,\rm{K}) + 1.537 \ 10^{-6} \times (T-300\,\rm{K})^2\right)
\end{cases}
\end{equation}

This model is deduced from ref.~[\onlinecite{Segura2020}], by using an interpolation valid over the $273-500$~K temperature range. All energy units are in reciprocal centimeters.

We compute a  $+1.4\%$ absolute absorptance change in this case Fig.~\ref{F160_thermochroism}B), thus supporting the idea that hBN is responsible for the slight change in absorptance. In any case, \eb{as}{} this change is extremely small, \eb{}{demonstrating that even at MIR photon energies, electro- or thermochroism do not play a significant part in the absorptance spectrum of our GFETs. Thus,} we can generally conclude that the radiated power change with bias is solely due to graphene's electron and hole population statistics.

\subsubsection{The contribution of intraband and interband transitions to the emission}

\begin{figure}[h!]
\hspace*{-1cm}
\centering{}\includegraphics[width=12cm]{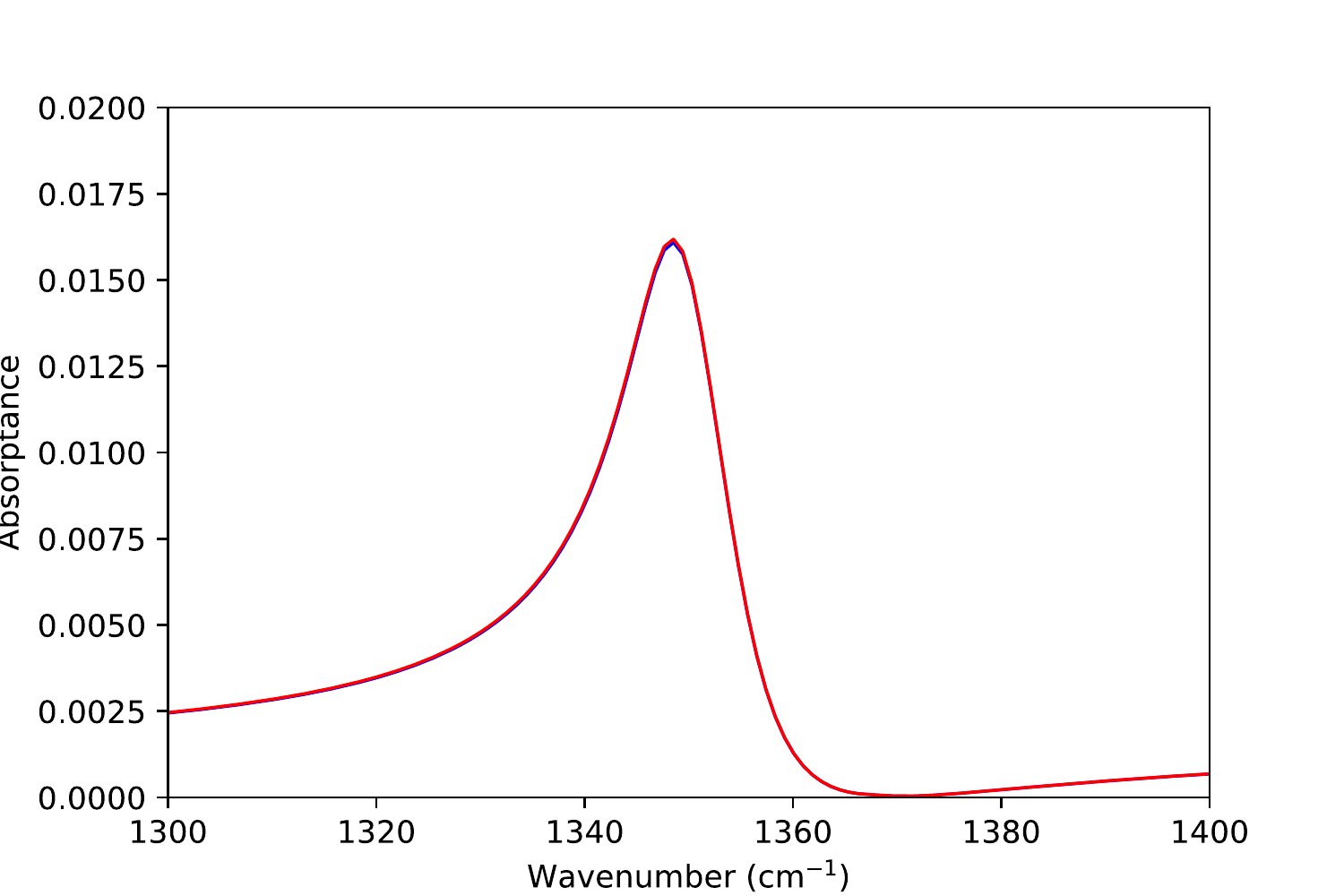}
\caption{Intraband versus Interband contribution to emittance. By increasing the carrier lifetime of graphene's electrons in the optical conductivity formula of graphene, we can suppress the role of intaband transitions in the calculation of light-matter coupling. Here, only the absorptance of graphene within the heterostructure is represented for the two configurations: with intraband and interband transitions (blue) and with interband transitions only (red). The integrated difference over the peak width accounts for only $-0.3\%$. ($E_{F} = 30\, \mathrm{meV}, T = 300 \, \mathrm{K}$)}
\label{F160_intra_inter}
\end{figure}

Finally, as LKL relies on the assumption that only direct optical transitions are involved in the light emission process, we have to verify this assumption for graphene. 
To this end, we utilize the decomposition of graphene's optical conductivity $\sigma=\sigma_{intra}+\sigma_{inter}$ introduced in Eq.~\ref{eq:sigma_intra_inter}. Since the optical conductivity directly shapes the absorptance of graphene within the heterostructure, we compare the computation of absorptance with and without the intraband contribution. The result of this comparison is presented in Fig.~\ref{F160_intra_inter}, in which the contribution of interband transitions is clearly found to be predominant in the far-field emission in the investigated MIR spectral range.

\subsubsection{Determination of radiative fluxes}

In this subsection, we explain how to obtain the chemical potential imbalance reported in section III from the comparison of the measured spectral flux and of the computed incandescent spectral flux. According to the Kirchhoff law (\ref{eq:GKL}), and given the absence of electro- and thermochroism, they both scale with the factor 
\begin{equation}
F(T_e, \Delta \mu) = 
\frac{1}{\exp\left(\frac{\hbar\omega-\Delta \mu}{k_{B}T_e}\right)-1} -
\frac{1}{\exp\left(\frac{\hbar\omega}{k_{B}T_0}\right)-1},
\label{eq:Boost}
\end{equation}

where the first term is the out-going radiative spectral flux, and the second term is the incoming radiative spectral flux due to the room-temperature $T_0$ environment of the transistor. 

The measured spectral flux is scaled with respect to the expected incandescent emission at the electronic temperature, that is $P_{\rm{rad}} = R\, P_{\rm{incan}},$ where  $P_{\rm{incan}} \propto F(T_e, 0),$ and, from the LKL $P_{\rm{rad}} \propto F(T_e, \Delta \mu).$ From these equations we determine the chemical potential imbalance $\Delta \mu$ as the solution of equation: 

\begin{equation}
F(T_e, \Delta \mu) = R\, F(T_e, 0).
\end{equation}

Likewise we obtain the effective radiation temperature $T_e^*$ from the equation

\begin{equation}
F(T_e^*, 0) = R\, F(T_e, 0).
\end{equation}